\documentclass[twocolumn,preprintnumbers,amsmath,amssymb,pre,floatfix,showkeys,superscriptaddress,nofootinbib,secnumarabic]{revtex4}

\usepackage[english]{babel}
\usepackage[latin2]{inputenc}
\usepackage[T1]{fontenc}

\usepackage{color}
\usepackage{amsmath}
\usepackage{amssymb}
\usepackage{ae,aecompl}
\usepackage{enumerate}
\usepackage{epsfig}
\usepackage{graphics}
\usepackage{graphicx}
\usepackage{pifont}
\usepackage[sort&compress]{natbib}
\usepackage{wasysym}
\usepackage{rotating}
\usepackage{multirow}

\def\ev #1{\left\langle #1 \right\rangle}
\def\evl #1{\overline{#1}}

\newcommand{\addfig}[2]
{
\begin{figure}[ptb]
\centerline{\includegraphics[height=210pt]{#1}}
\caption{#2}
\label{fig:#1}
\end{figure}
}

\newcommand{\addwidefig}[2]
{
\begin{figure*}[ptb]
\centerline{\includegraphics[width=360pt]{#1}}
\caption{#2}
\label{fig:#1}
\end{figure*}
}

\newcommand{\addverywidefig}[2]
{
\begin{figure*}[ptb]
\centerline{\includegraphics[width=550pt]{#1}}
\caption{#2}
\label{fig:#1}
\end{figure*}
}

\newcommand{\addtwofigs}[3]
{
\begin{figure*}[ptb]
\centerline{\includegraphics[height=190pt]{#1}\includegraphics[height=190pt]{#2}}
\caption{#3}
\label{fig:#1}
\end{figure*}
}

\def\etal{{\it et al}. }
\def\ccc#1;#2{\left\langle #1 \left\vert #2 \right.\right\rangle}
\def\ev #1{\left\langle #1 \right\rangle}

\begin{document}

\preprint{}
\title{Fluctuation scaling in complex systems: Taylor's law and beyond}
\author{Zolt\'an Eisler}
\email{eisler@maxwell.phy.bme.hu}
\affiliation{Science \& Finance, Capital Fund Management, Paris, France}
\affiliation{Department of Theoretical
Physics, Budapest University of Technology and Economics, Budapest, Hungary}
\author{Imre Bartos}
\affiliation{Department of Physics of Complex Systems, Lor\'and E\"otv\"os University, Budapest, Hungary}
\affiliation{Columbia University, Department of Physics, New York, USA}
\author{J\'anos Kert\'esz}
\affiliation{Department of Theoretical
Physics, Budapest University of Technology and Economics, Budapest, Hungary}
\affiliation{Physics of Condensed Matter Group, HAS, BME, Budapest, Hungary}

\date{\today}



\begin{abstract}
Complex systems consist of many interacting elements which participate in some dynamical process. The activity of various elements is often different and the fluctuation in the activity of an element grows monotonically with the average activity. This relationship is often of the form "$\mathrm{fluctuations} \approx \mathrm{const.}\times \mathrm{average}^\alpha$", where the exponent $\alpha$ is predominantly in the range $[1/2, 1]$. This power law has been observed in a very wide range of disciplines, ranging from population dynamics through the Internet to the stock market and it is often treated under the names \emph{Taylor's law} or \emph{fluctuation scaling}. This review attempts to show how general the above scaling relationship is by surveying the literature, as well as by reporting some new empirical data and model calculations. We also show some basic principles that can underlie the generality of the phenomenon. This is followed by a mean-field framework based on sums of random variables. In this context the emergence of fluctuation scaling is equivalent to some corresponding limit theorems. In certain physical systems fluctuation scaling can be related to finite size scaling.

\end{abstract}

\maketitle

\vskip1cm

\rightline{Dedicated to the memory of}
\rightline{L.R. Taylor (1924--2007)}

\tableofcontents



\section{Introduction}
\label{toc:1}
Interacting systems of many units with emergent collective behavior are often termed "complex".
Such complex systems are ubiquitous in many fields of research ranging from engineering sciences through physics and biology to sociology. An advantage of the related multi-disciplinary approach is that the universal appearance of several phenomena can be revealed more easily. Such generally observed characteristics include (multi-)frac\-tality or scale invariance \cite{vicsek.book, bak.soc}, the related Pareto or Zipf laws \cite{pareto,zipf}, self-organized and critical behavior.


In this paper we study such a general feature related to the scaling properties of the fluctuations in complex systems. This type of scaling relationship is called Taylor's law by ecologists after L.R. Taylor and his influential paper on natural populations in 1961 \cite{taylor}. The law states that for any fixed species the fluctuations in the size of a population (characterized by the standard deviation) can be approximately written as a constant times the average population to a power $\alpha$: $$\mathrm{fluctuations} \approx \mathrm{const.}\times \mathrm{average}^\alpha$$ for a wide range of the average.

The phenomenon was -- to our knowledge -- first discovered in 1938 by H. Fairfield Smith \cite{fairfield.smith}, who wrote an equivalent formula for the yield of crop fields though his paper has, surprisingly, received much less attention than Taylor's work. {The same relationship was explored recently by Menezes and Barab\'asi \cite{barabasi.fluct} for dynamics on complex networks, and later termed "fluctuation scaling" \cite{eisler.unified} in the physics literature. There the temporal fluctuations and the averages of the network's traffic were measured at the different nodes.}

Despite the analogous questions, the exchange of ideas between disciplines is very limited. This review attempts to show how general fluctuation scaling is by surveying the literature and the current models, as well as by reporting some new empirical evidence and presenting new model calculations. We also have the aim to step beyond mere demonstration, and show some basic principles that can potentially underlie the generality of the phenomenon. 

The paper is organized as follows. Section \ref{sec:fs} gives a more precise definition of fluctuation scaling. We then give a brief overview of empirical results from the literature, and also some previously unpublished findings.\footnote{Please note that "$\log$" will always denote $10$-base logarithms.} Section \ref{sec:gen} presents a general mean-field formalism based on sums of random variables. This is followed by the interpretation of the scaling exponent $\alpha$, and how it reflects the dynamics of the complex system. We show that $\alpha$ is usually between $1/2$ and $1$, and that these two limiting values can both arise from several, simple types of dynamics. We then show three scenarios how intermediate exponents can arise. Because of this multitude of possibilities no value of $\alpha$ can be used \emph{in itself} to uniquely identify the internal dynamics of a system, but it is still possible to exclude many options that would be incompatible with the observed value of $\alpha$. The remaining possibilities can be
narrowed down further by analyzing the time scale dependence of $\alpha$ and by the application of our mean field framework. The procedure is demonstrated on some simple models in Section \ref{sec:mod}, and the relationship between fluctuation scaling, (self-organized) criticality, scaling and multiscaling is explored. Fluctuation scaling can be directly applied to certain physical systems, where one finds a strong connection with finite size scaling. Section \ref{sec:disc} gives a general discussion and directions for future research. Finally Section \ref{sec:conc} concludes. Some calculations were deferred to the Appendices.

\section{Fluctuation scaling}
\label{sec:fs}
\label{toc:2}

\subsection{Basic notions}
\label{toc:2.1}

Throughout the paper we will always consider some additive quantity $f$, and the dependence between its mean and standard deviation. By dependence we mean the behavior of $f$ over a multitude of observations. Say, if we can observe the same dynamical variable in several settings where it has different means, how does the standard deviation change with the value of the mean?

In order to determine this dependence one needs many realizations. These can be simultaneous temporal observations for different elements (nodes, subsystems) of a large complex system. The measured means and standard deviations are then calculated in time, and the subsystems are compared: for subsystems with a larger mean $f$ are the fluctuations larger as well?

{In other cases $f$ is \emph{not} considered as time dependent, only as a fixed value for every subsystem.  Then the averages are taken over an ensemble of subsystems of equal size, and the standard deviation characterizes the variation of $f$ between subsystems of the same size.}

We just used the expressions "elements", "nodes", "subsystems" and "the same size", but what do these mean? Imagine that we want to quantify fluctuations in the traffic of Internet routers. It is very straightforward to calculate the mean and the standard deviation of, say, daily data throughput, and the question whether routers with larger mean traffic exhibit larger fluctuations can be investigated. However, routers are not "subsystems" of the Internet. They merely represent \emph{points of measurement}, elements of the system. The traffic is formed as a sum of data packets that are "extrinsic" to the elements. The packets do not belong to the structure of the network, but they carry the dynamics on it.
Here we are not interested in the structure of the routers, i.e., the nodes. Instead, the data over which the averages are taken have a \emph{temporal} structure.

Let us take a different example. Now we want to analyze data on the size of animal populations. A population can be divided into smaller groups, which then consist of individuals, and this gives a true notion of size. Various smaller areas can be naively considered as "subsystems" with respect to the habitat of the species, for example a continent. These subsystems are \emph{not structureless}, and their population comes about as a sum over their smaller subgroups of individuals.

In our review we will call the points of measurement as \emph{nodes}, whether they have a structure or not. The additive quantity under study, be it activity, population, traffic or whatever else, will be denoted by $f_i$, where $i$ indicates the node of measurement. This will always be decomposed as a sum of random variables, which will either represent internal constituents or some events similar to the arrival of the extrinsic "packets" to the Internet routers. In both cases we will call these the \emph{constituents} of the nodes/signals.  Their number for node $i$ will be denoted by $N_i$, and their respective contributions to $f$ will be denoted by $V_{i,n}$, where $n=1\dots N_i$. Examples of the scheme for building up a system from constituents can be seen in Table  
\ref{tab:scheme}. Now we turn to more precise definitions.

\begin{table*}[!ht]
\begin{align}
	\mathrm{constituents} \rightarrow \mathrm{nodes/elements} \rightarrow \mathrm{complex\ system} \nonumber \\
	V_{i,n}, n=1,\dots,N_i \rightarrow f_i, i = 1,\dots,M \rightarrow \mathrm{total\ activity} \nonumber \\
	\mathrm{a\ group\ of\ individuals} \rightarrow \mathrm{a\ population} \rightarrow \mathrm{a\ species} \nonumber \\
	\mathrm{a\ single\ tree} \rightarrow \mathrm{a\ single\ forest} \rightarrow \mathrm{all\ forests\ of\ a\ continent} \nonumber \\
	\mathrm{a\ single\ data\ packet} \rightarrow \mathrm{router} \rightarrow \mathrm{Internet}\nonumber \\
	\mathrm{a\ single\ car} \rightarrow \mathrm{measurement\ point} \rightarrow \mathrm{highway\ system}\nonumber 
	\end{align}
	\caption{Examples of building up a system from its constituents. The symbol $\rightarrow$ indicates that the kind of object on the right is made up by several ones of the kind on the left.}
	\label{tab:scheme}
\end{table*}

\subsubsection{Temporal fluctuation scaling (TFS)}
\label{toc:2.1.1}
Let us assume that during an extended period we can measure an additive quantity $f_i$ within a system at its nodes (labeled by the index $i$). For some finite time duration $[t, t+\Delta t)$ the signal can be formally decomposed as the sum
\begin{equation}
	f_i^{\Delta t}(t)=\sum_{n=1}^{N_i^{\Delta t}(t)} V_{i,n}^{\Delta t}(t).
	\label{eq:sum}
\end{equation}
$N_i^{\Delta t}(t)$ is the number of constituents of node $i$ during $[t,t+\Delta t)$. We assume that $V_{i,n}^{\Delta t}(t)\geq 0$, so that the time average of $f_i^{\Delta t}$ does not vanish. For example, if on the stock market during $[t, t+\Delta t)$ there are $N_i^{\Delta t}(t)$ transactions with the papers of the $i$'th company, and the $n$'th of those transactions has a value $V_{i,n}^{\Delta t}(t)$, then the total trading activity of stock $i$ can be calculated by this formula.

The time average of Eq. \eqref{eq:sum}, which we will denote as $\ev{f_i^{\Delta t}}$, can be calculated as
\begin{equation}
	\ev{f_i^{\Delta t}}=\frac{1}{Q}\sum_{q=0}^{Q-1} f_i^{\Delta t}(q\Delta t) = \frac{1}{Q}\sum_{q=0}^{Q-1} \sum_{n=1}^{N_i^{\Delta t}(q\Delta t)} V_{i,n}^{\Delta t}(q\Delta t),
	\label{eq:timeavg}
\end{equation}
where $Q=T/\Delta t$, and $T$ is the total time of measurement. From the definitions it is trivial that $\ev{f_i^{\Delta t}} = \Delta t \ev{f_i^{\Delta t=1}}$. We will use $\ev{f_i}$ without the upper index to denote this latter quantity.

On any time scale the variance can be obtained as a time average:
$$\sigma^2_i(\Delta t) = \ev{[f_i^{\Delta t}]^2}- \ev{f_i^{\Delta t}}^2,$$
this quantity characterizes the fluctuations of the activity of a fixed node $i$ from interval to interval.

When $f$ is positive and additive, it is often observed that the relationship between the standard deviation and the mean of $f$ is given by a power law:
\begin{equation*}
	\sigma_i(\Delta t) \propto \ev{f_i^{\Delta t}}^{\alpha_\mathrm{T}},
\end{equation*}
where one varies the node $i$, and $\Delta t$ is fixed. The dependence of the right hand side on $\Delta t$ is trivial, since $\ev{f_i^{\Delta t}}\equiv \Delta t \ev{f_i}$. Thus throughout the paper we will use $\ev{f_i}$ as the scaling variable:
\begin{equation}
	\sigma_i(\Delta t) \propto \ev{f_i}^{\alpha_\mathrm{T}}.
	\label{eq:taylori}
\end{equation}
The exponent $\alpha_\mathrm{T}$ is usually in the range $[1/2,1]$. The lower index $T$ in the scaling exponent indicates that the statistical quantities are defined as \emph{temporal} averages as in Eq. \eqref{eq:timeavg}.

Finally, if the $i$-dependence of $\sigma$ and $\ev{f}$ is only manifested via a well defined parameter of the nodes, such as their linear extent ($L$), area ($A$), a fixed number of constituents ($N$) or some other size-like parameter $S$, then we will use this quantity as lower index where possible. For example temporal standard deviation will be denoted as $\sigma_S$.

\subsubsection{Ensemble fluctuation scaling (EFS)}
\label{toc:2.1.2}
Again imagine that nodes have a well defined size-like parameter $S$, and it is possible to group them according to that. Furthermore, assume that nodes that fall into the same group have equivalent statistical properties. Then aside from the temporal average given separately for each node, one can also define the average of $f$ \emph{within each group}. This is a sort of ensemble average over similar nodes, it will be denoted by $\evl{f_S^{\Delta t}}$, and it can be calculated as
\begin{equation}
	\evl{f_S^{\Delta t}}=\frac{1}{M_S}\sum_{\forall i: S_i = S} f_i^{\Delta t}(t)
\end{equation}
Both $t$ and $\Delta t$ are now fixed, the summation instead goes for those nodes $i$ which have a size $S_i=S$, and $M_S$ is the number of such nodes. In the notation we will omit $t$ for simplicity. Variance is given by
$$\evl{\sigma^2_S}(\Delta t) = \evl{[f_S^{\Delta t}]^2}- \evl{f_S^{\Delta t}}^2.$$

Fluctuation scaling can also arise here in the form
\begin{equation}
	\overline{\sigma_S}(\Delta t) \propto {\overline{f_S}}^{\alpha_\mathrm{E}},
	\label{eq:taylorN}
\end{equation}
where we compare different groups by varying $S$, while $\Delta t$ is kept constant. For convenience we follow the convention of the previous part: On the right hand side of Eq. \eqref{eq:taylorN} we write $\evl{f_S}$, which is a short notation for $\evl{f_S^{\Delta t = 1}}$. The scaling exponent $\alpha_\mathrm{E}$ will always indicate when we use \emph{ensemble} averaging over elements of the same size.

For data analysis the size $S$ very often corresponds to the linear size $L$, or the area $A$ of the node/subsystem, and there the lower index will be changed accordingly. For example, the classic study of Taylor \cite{taylor} compares areas of different size $A$, and the measured quantity is the population size of a given species in the area. The constituents can be smaller groups or, as usually called, metapopulations. If one considers the number of groups ($N$) and the size of the groups ($V_n$) as random variables, the total population has the same sum form as before:
\begin{equation*}
		f_i=\sum_{n=1}^{N_i} V_{i,n},
\end{equation*}
which is the analogue of Eq. \eqref{eq:sum}.

We will call the relationship \eqref{eq:taylori} temporal fluctuation scaling (TFS), and \eqref{eq:taylorN} ensemble fluctuation scaling (EFS). When we do not wish to distinguish between the two cases, we will simply use fluctuation scaling (FS), and then the exponent will be denoted by $\alpha$ without lower index. There exists a large body of results on these subjects, and the literature is spread over many disciplines. Therefore in the following we would like to give a (necessarily incomplete) overview of the results. A summary is presented in Table \ref{tab:examples}.

\subsection{Empirical results: ensemble averages}
\label{toc:2.2}
\subsubsection{Pioneering studies}
\label{toc:2.2.1}
As noted in the introduction, the first observations of fluctuation scaling appeared in two independent studies, well before the wide\-spread recognition of fractality and scaling \cite{mandelbrot.book}. The paper of Fairfield Smith \cite{fairfield.smith} was published in 1938, and it was concerned with the yields of agricultural crops. For a fixed size of land ($A$) it is possible to calculate the average yield $\evl{f_A}$ of a certain type of crop, and the standard deviation $\evl{\sigma_A}$ of the yield between areas of size $A$. Then the calculation can be done for areas of many different sizes. It was found that there exists the power law \eqref{eq:taylorN} relationship between the two quantities, $\evl{\sigma_A} \propto {\evl{f_A}}^{\alpha_\mathrm{E}}$, with $\alpha_\mathrm{E} \approx 0.62$.

Taylor's 1961 paper \cite{taylor} stated the scaling law \eqref{eq:taylorN} for systems in population dynamics. Similarly to Fairfield Smith, Taylor took an ensemble of areas of the same size, and measured the number of individuals of a certain type of animal. With increasing area size both the mean and the variance of the population grew, with a power law relationship between the two quantities. Let us now take a closer look at fluctuations in ecology.

\subsubsection{Ecology}
\label{toc:2.2.2}
Stable populations in a given habitat fluctuate around a typical size called the habitat's carrying capacity \cite{maurer.taper, sheep, noisy.clockwork}. These fluctuations have a very rich internal structure \cite{noisy.clockwork}. Both the randomness of birth-death processes (a kind of "intrinsic noise") and external climatic forcing play an important role \cite{noisy.clockwork, saether}. The effect of climatic factors is so strong that it can synchronize the fluctuations of even non-interacting populations (the so-called Moran effect) \cite{sheep, moran}. To further complicate the situation, individuals of a species interact among themselves, just as well as species interact with each other. These interactions are non-linear and by now they are also commonly recognized to have a significant dependence on the population density/size. So interactions, driving and noise all contribute to population dynamics to a certain degree \cite{noisy.clockwork, saether}. This diversity makes any ecosystem a showcase of complexity; certain regularities are known, but the bigger picture is still missing.

This is the reason why by discovering a universal law, Taylor's paper \cite{taylor} triggered a growing activity in ecology, with literally a thousand publications to date. Taylor's results were verified for a wide range of populations, and the value of the exponent was predominantly found to be $1/2 \leq \alpha_\mathrm{E} \leq 1$ \cite{conserv}. Despite its generality, the origins of the law and the meaning of $\alpha$ are still much debated. Anderson \etal \cite{anderson.variability} suggested that the influence of environmental fluctuations may be responsible for the observed non-trivial exponents. The model of Kendal \cite{kendal.dla} proposed a dynamics similar to Diffusion Limited Aggregation in which self-similarity gives rise to the mean-variance scaling. Another study \cite{kendal.ecological} proposed that the exponents can be described by a class of statistical models, which rely on the interplay between the number of animal clusters in an area and the size of the individual clusters. We will discuss these models in detail in Section \ref{sec:tweedie}. For two comprehensive reviews of these (and more) scaling laws in ecological and related systems see Kendal \cite{kendal.ecological} and Marquet \etal \cite{marquet.scaling}.

\addfig{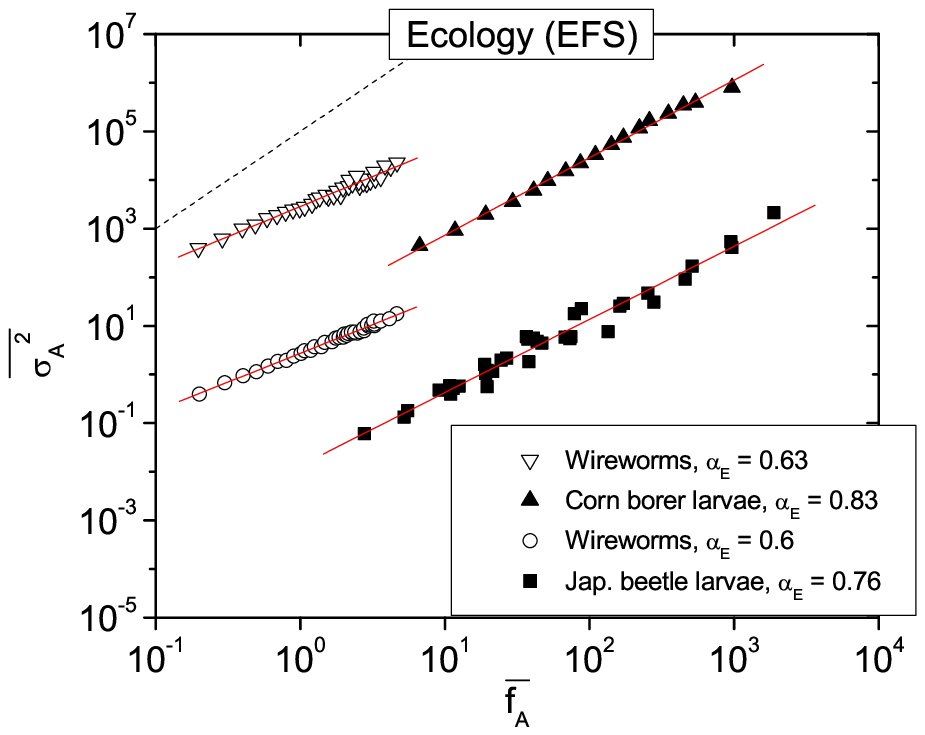}{Fluctuation scaling for ensemble averages of the population of four species. Every point represents the mean $\overline{f_A}$ and variance $\overline{\sigma^2_A}$ over an ensemble of areas of the same size $A$ \cite{taylor}. The bottom dashed line corresponds to $\alpha_E=1/2$, the top one to $\alpha_E=1$. Points were shifted both vertically and horizontally for better visibility.}

\subsubsection{Life sciences}
\label{toc:2.2.3}

There is a number of findings from cellular and molecular biology regarding FS. Azevedo and Leroi \cite{cells} conducted a very extensive study of how the typical cell count of a species is related to its fluctuations from individual to individual. They found that FS holds over almost $10$ orders of magnitude in size, between more than $2000$ species, see Fig. \ref{fig:Gcellcount}. The exponent $\alpha_\mathrm{E}$ differs among tissue types, but for entire organisms its value is approximately $1$. Kendal \cite{kendal.tumor} presents similar findings for the number of tumor cells in groups of mice, but the exponents vary. 

Similarly, Kendal \cite{kendal.genomevar} analyzed for the most common variations in the human genome called Single Nucleotide Polymorphisms (SNPs) \cite{genomevar}. He found that the mean and the variance of the number of SNPs in a DNA sequence scale as different non-trivial powers of the length of the sequence, and thus the variance also scales with the mean.

\addfig{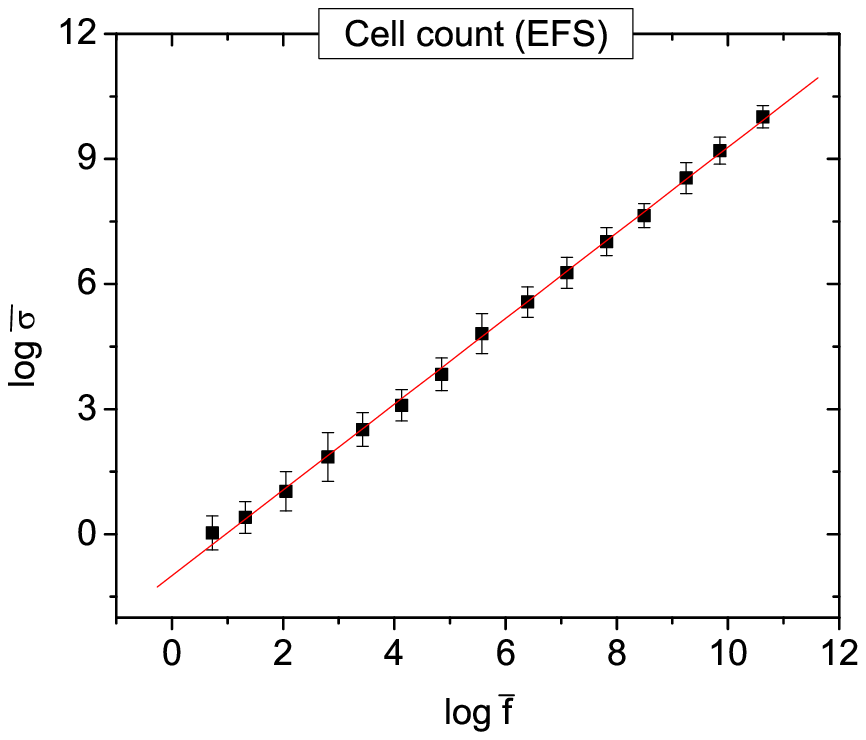}{Fluctuation scaling for the cell count of species. For every species the average cell count $\overline{f}$ and its variance $\overline{\sigma^2}$ was calculated separately. Then these points were binned logarithmically for better visibility, error bars show the standard deviation of $\log \overline \sigma$ in the bins. $\alpha_\mathrm{E} = 1$. Data courtesy of Ricardo Azevedo \cite{cells}.}

\subsubsection{Physics}
\label{toc:2.2.4}
FS has been present in the physics literature as well. Many extensive quantities are known to have equilibrium fluctuations proportional to the square root of the system size, implying the value $\alpha_\mathrm{E} = 1/2$ \cite{reichl, landau5}. This relationship is a simple consequence of the central limit theorem (see Section \ref{sec:alpha1/2}). Botet \etal \cite{botet.prl, botet.pre, botet.npb} find EFS for a wide range of models, and also for the fragment multiplicity measured in heavy-ion collision experiments. 
Moreover, a linear ($\alpha = 1$) relationship was found between the fluctuations and mean fluxes of cosmic radiation \cite{uttley.flux, vaughan.flux}. Here the ensemble is formed by cutting a single time series into pieces, and then periods with higher average activity exhibit higher fluctuations.

\subsection{Empirical results: temporal averages}
\label{toc:2.3}
\label{sec:empiricaltime}

We will now turn to temporal FS. For the collection of such data it is necessary to have multi-channel measurements, simultaneously monitoring the behavior of a range of elements $i$. With the unbroken growth of computing infrastructure, many technological networks now offer appropriate datasets, several ones publicly available.

\subsubsection{Complex networks}
\label{toc:2.3.1}
Menezes and Barab\'asi \cite{barabasi.fluct,barabasi.separating}, in part inspired by Taylor's original paper, found TFS for several complex networks. {A good example is the analysis of Internet traffic, which was later revisited by Duch and Arenas \cite{duch.internet}. In their study they analyzed the traffic of the Abilene backbone network. The nodes $i$ correspond to routers, and the mean and variance of their data flow was calculated. In Fig. \ref{fig:Ginternet1w} we show their results for weekly data traffic, the best fit is achieved with $\alpha_\mathrm{T} \approx 0.75$. Menezes and Barab\'asi also analyzed web page visitations, river flow, microchip logical gates and highway traffic. They proposed that the datasets should fall into two "universality classes" with $\alpha_\mathrm{T} = 1/2$ and $1$. There also exists a growing body of literature on transport processes on networks, and the scaling of fluctuations in such systems \cite{eisler.internal, duch.internet, menezes.flux, tadic.loops, duch.model}.}

\addfig{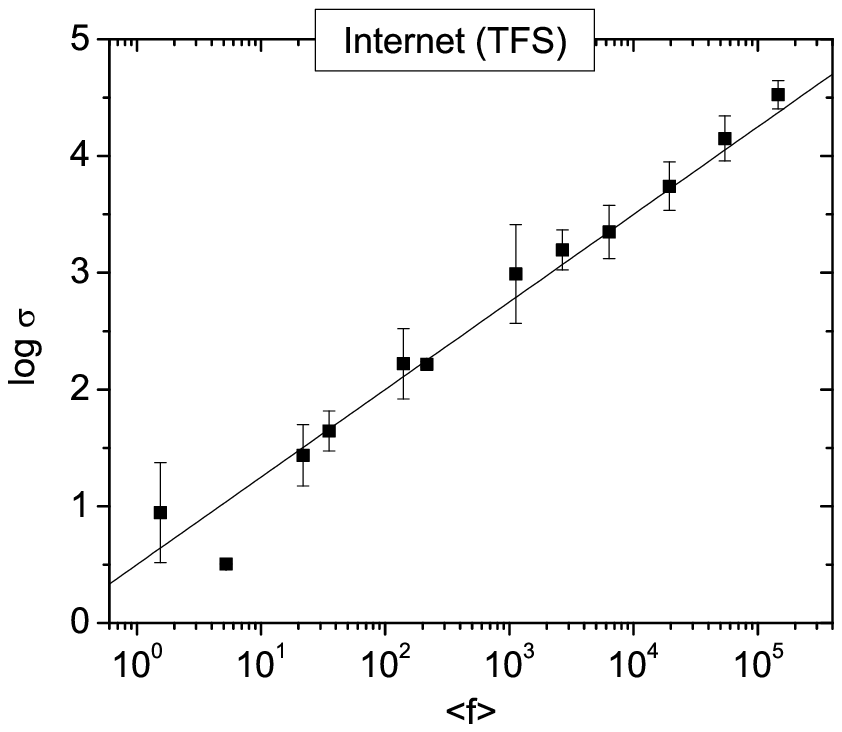}{Fluctuation scaling for the traffic of Internet routers. For every router the temporal average of weekly data traffic $\ev{f_i}$ and its standard deviation $\sigma_i$ was calculated separately. Then these points were binned logarithmically for better visibility, error bars show the standard deviation of $\log \sigma$ in the bins. The fitted exponent is $\alpha_\mathrm{T} = 0.75$. Data courtesy of Jordi Duch and Alex Arenas \cite{duch.internet}.}

\subsubsection{Ecology}
\label{sec:timeecol}
\label{toc:2.3.2}
{Ecologists have made many advances regarding TFS as well, but the literature is far from unequivocal. The basic concept is to monitor many populations of a given species for an extended period of time, and then for each population $i$ calculate the temporal mean $\ev{f_i}$ and standard deviation $\sigma_i$ of abundance. These are typically power law related according to TFS, examples are shown in Fig. \ref{fig:Ginsect}.}

\addfig{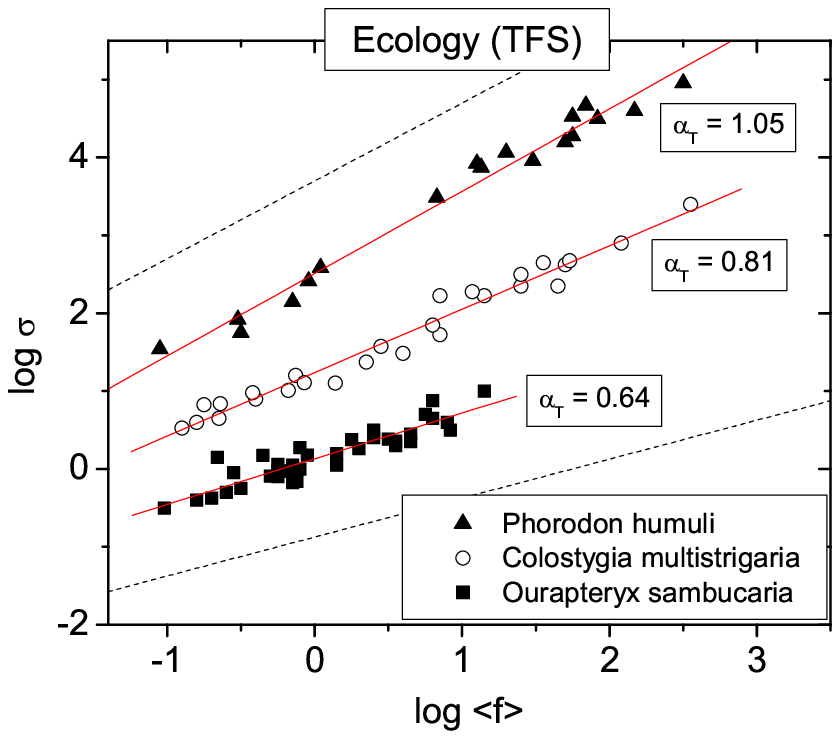}{Fluctuation scaling for temporal averages of the population of three species. A point represents the temporal mean $\ev{f_i}$ and standard deviation $\sigma_i$ of a population. The bottom dashed line corresponds to $\alpha_T=1/2$, the top one to $\alpha_T=1$. Points were shifted both vertically for better visibility. Data courtesy of Marm Kilpatrick \cite{kilpatrick.ives,taylor.insect}.}

Classical population dynamics offers several benchmark models \cite{may.complexity, lotka, volterra}, but simple deterministic and Markovian models cannot explain the observed $\alpha_\mathrm{T}$ values between $1/2$ and $1$. After a range of small populations where they show realistic behavior, they cross over to either $\alpha_\mathrm{T} = 1/2$ or $1$ \cite{keeling.simplestoch}. The model of Kilpatrick and Ives \cite{kilpatrick.ives} suggested that the interaction between species and feedback mechanisms between their fluctuations can give rise to any value of $\alpha_\mathrm{T}$. Perry proposed an even simpler chaotic model \cite{perry.chaotic}. Both of these models can yield various exponents, but still only when populations are small enough.

There have been several findings for plant species. In a series of papers Ballantyne and Kerkhoff showed that the reproductive (yearly seed count) variability of trees follows TFS with $\alpha_\mathrm{T} = 1$. The same value is supported by the Satake-Iwasa \cite{satake.iwasa} forest model. There the trees are modeled by interacting oscillators which synchronize above a critical value of the coupling \cite{ballantyne.model, ballantyne.correls}. The synchronization transition coincides with a transition from $\alpha_\mathrm{T} = 1/2$ to $\alpha_\mathrm{T} = 1$. \footnote{Similar synchronization mechanism has also been observed in the reproduction of animals \cite{sheep}.} 

We will now briefly describe their empirical study \cite{ballantyne.scaling}. The dataset \cite{koenig.masting} consists of yearly observations of the seed production of trees throughout the Northern Hemisphere. In particular, we considered three subsets of the dataset \cite{uponrequest}, those collected by Tallqvist \cite{tallqvist}, Franklin \cite{franklin} and Weaver and Forcella \cite{weaver.forcella}, including $4-17$ years of observations for $44$, $148$ and $28$ sites, respectively. The fits for TFS are given in Fig. \ref{fig:Gtreesalpha}(left). The exponents for the three subsets were found to be $\alpha_\mathrm{T} = 0.97$, $0.93$ and $0.90$. Given the quality of the fits it is not possible to outrule that for all three datasets $\alpha_\mathrm{T} = 1$ (as suggested by Ref. \cite{ballantyne.scaling}). However, here we make an attempt to give an argument that predicts otherwise and can be tested.

\addtwofigs{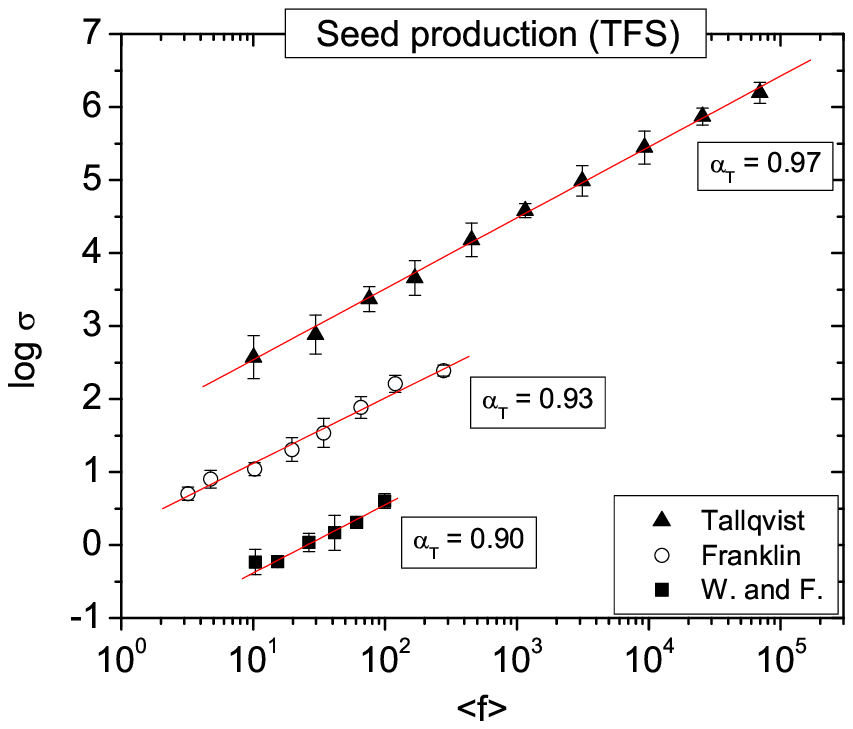}{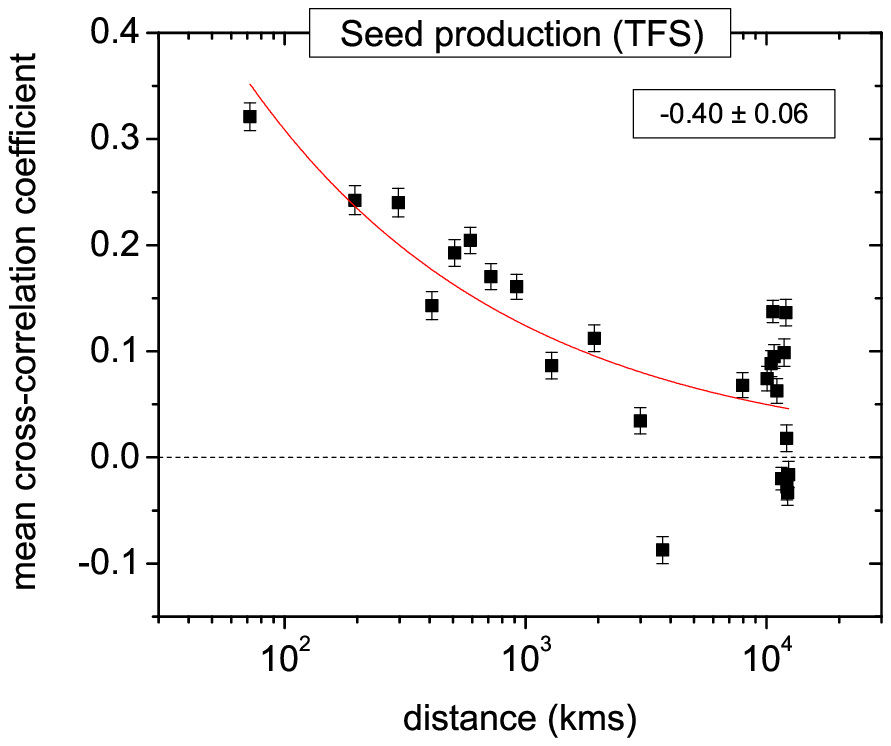}{(left) Fluctuation scaling for the yearly seed count (reproductive activity) of trees from three studies. The fitted exponents are $\alpha_\mathrm{T} = 0.90, 0.93, 0.97$. Points were logarithmically binned and $\log \sigma$ was averaged for better visibility, error bars show the standard deviation of $\log \sigma$ in the bins. The estimates are close to, but below $1$. (right) The average cross-correlation coefficient between pairs of sites as a function of their distance. Pairs of sites were binned to have an equal number per bin, error bars show the standard deviation of the cross correlation coefficients in the bins. The red line is a fit with the power law $C(r) \propto r^{-0.40}$, from the fit the error of the exponent is $\pm 0.06$.}

Simulations of the Satake-Iwasa model already suggested that long-range synchronization can cause $\alpha_\mathrm{T} > 1/2$, and the presence of such a tendency is well known for trees. Koenig and Knops \cite{koenig.masting} conclude that there exists a significant positive correlation between the reproductive activity of trees for distances longer than $1000$ kms (this phenomenon is called masting in the ecology literature). While Ref. \cite{koenig.masting} is much more precise and detailed, we also outline a simple measurement: in Fig. \ref{fig:Gtreesalpha}(right) we plot the average $C(r)$ cross-correlation coefficients between the sites in the complete dataset as a function of the distance $r$ of the sites. As expected, we find that cross-correlations decay very slowly with distance, and  the dependence can be fitted approximately by
\begin{equation}
	C(r) \propto r^{-0.40},
\end{equation}
(although admittedly the fit is not perfect).
Sections \ref{sec:sync} and \ref{sec:binary} will show, that while perfect synchronization a'la Satake-Iwasa leads to $\alpha_\mathrm{T} = 1$, partial synchronization with the above power-law correlations, implies $\alpha_\mathrm{T} = 1-0.40/2 = 0.8$, see 
Eq. \eqref{eq:popalpha}. A better quantitative agreement would warrant larger datasets which are not available at present, but there have been some promising attempts along the same lines \cite{ballantyne.correls}.

\subsubsection{Life sciences}
\label{toc:2.3.3}

Keeling and Grenfell \cite{measles} suggested TFS for the size of epidemics, and found both empirically and by a simple Markov chain model of population dynamics that vaccination in general decreases not only the size of epidemics but also the value of $\alpha_\mathrm{T}$. TFS was later found by Woolhouse \etal \cite{pathogen} to also hold between different pathogens. TFS has been found in the cell-to-cell variation of protein transcription by Bar-Even \etal \cite{bareven.protein}, albeit with a crossover and a rather narrow range.

\subsubsection{Stock market}
\label{sec:stock}
\label{toc:2.3.4}

In this section we summarize the results of a series of papers \cite{eisler.non-universality, eisler.unified, eisler.sizematters, eisler.sizematters2}. The work was based on a TAQ database \cite{taq2000-2002}, recording the transactions of the New York
Stock Exchange (NYSE) for the years $2000-2002$. Very similar results were obtained for the NASDAQ and Chinese markets \cite{jiang.fluxes}.

We define the activity of stock $i$ as its total traded value, given as
$$f_i^{\Delta t}(t) = \sum_{n=1}^{N^{\Delta t}_i(t)} V^{\Delta t}_{i,n}(t),$$ where $N^{\Delta t}_i(t)$ is the number of transactions of stock $i$ in the period $[t, t+\Delta t)$. The individual values of these transactions are denoted by $V^{\Delta t}_{i,n}(t)$. Data were detrended by the well-known $U$-shaped daily pattern of traded volumes \cite{eisler.non-universality}.

Then the measurement of mean and variance was carried out. The exponent $\alpha_\mathrm{T}$ shows a strong dependence on the window size $\Delta t$, we will return to this result in Section \ref{sec:gamma}. The values range between $\alpha_\mathrm{T} = 0.68-0.87$, see Fig. \ref{fig:Lstock}.

\addfig{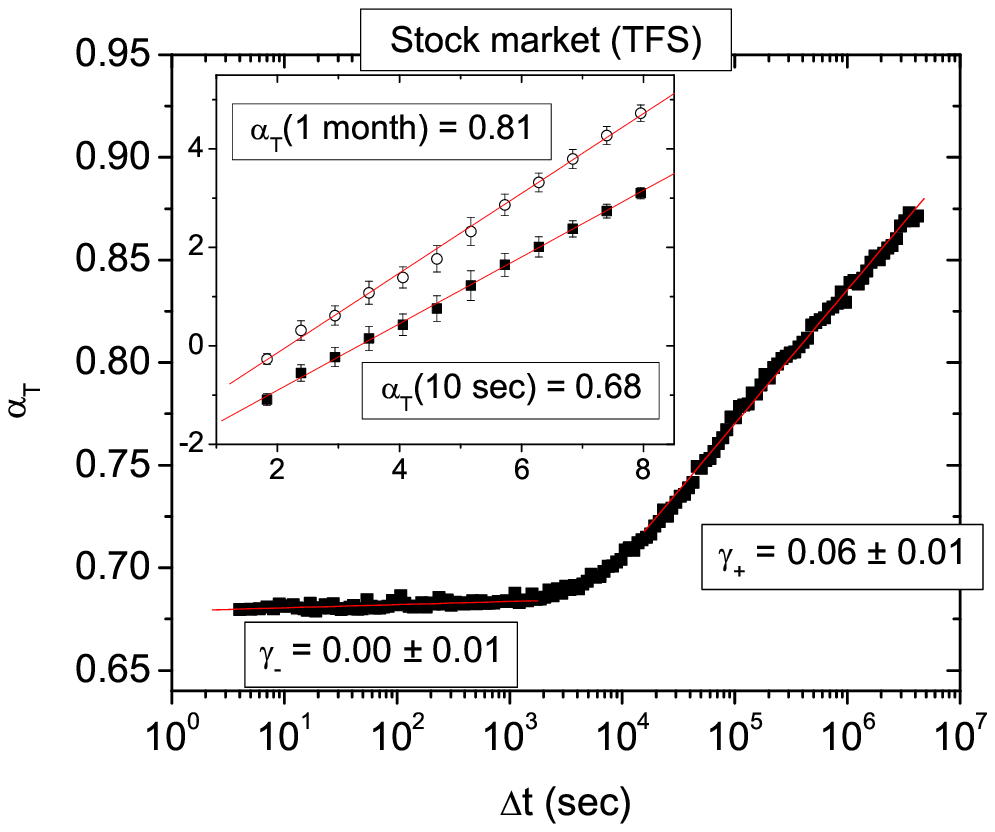}{The dependence of the FS exponent in stock market data on the size of the time window $\Delta t$. The dependence is logarithmic in two regimes, with the coefficients $\gamma_- \approx 0.00$ for $\Delta t < 10^3$ sec and $\gamma_+ \approx 0.06$ for $\Delta t > 3\times10^4$ sec. The crossover regime corresponds to the time scale of one trading day. Inset: FS for the fluctuations of the traded value of stocks for the window sizes $\Delta t = 10$ sec and $\Delta t = 1$ month. Points were logarithmically binned and $\log \sigma$ was averaged for better visibility, error bars show the standard deviation of $\log \sigma$ in the bins.}

When $\Delta t$ is very small, Ref. \cite{eisler.unified} shows that the individual transactions can be treated as independent events. Moreover, for the large enough stocks the average size of transactions ($\ev{V_i}$) can be calculated as a power of the mean number of transactions as $$\ev{V_i}\propto\ev{N_i}^\beta,$$ with $\beta \approx 0.65$ \cite{eisler.unified}. Equivalents of this property recur in several FS-related contexts. We will devote Sections \ref{sec:beta}-\ref{sec:tweedie} to this observation, which we will call impact inhomogeneity \cite{eisler.unified}. We will also show how to map the value of $\beta$ onto non-trivial $\alpha_\mathrm{T}$ values. By that method [see Eq. \eqref{eq:alphabeta}] the corresponding $\alpha_\mathrm{T}$ value should be $0.70$, which is very close to the actual value $\alpha_\mathrm{T}(\Delta t \rightarrow 0) = 0.69$. 

Another general observation \cite{eisler.unified} is that if $\alpha_\mathrm{T}$ is a function of $\Delta t$, then FS can only hold, if this dependence is logarithmic (cf. Section \ref{sec:gamma}). For the stock market this is true in two distinct regimes and those are separated by a crossover. For $\Delta t < 10^3$ sec ($-$ sign) and $\Delta t > 3\times10^4$ sec ($+$ sign) one finds
$$\alpha_{T,\pm}(\Delta t) = \alpha^*_{T,\pm} + \gamma_\pm \log \Delta t,$$ 
with $\gamma_- \approx 0.00$, and $\gamma_+ \approx 0.06$. On the other hand, the Hurst exponent $H_i$ can be defined as \cite{vicsek.book,dfa} 
\begin{equation}
\sigma_i(\Delta t) = \ev{\left [ f_i^{\Delta t}(t) - \ev{f_i^{\Delta t}(t)}
\right ]^2}^{1/2} \propto \Delta t^{H_i}. \label{eq:hurst}
\end{equation}
For NYSE this equation is found to be valid with
$$H_{i;\pm} = H^*_\pm + \gamma_\pm \log \ev{f_i}.$$ Lower indices indicate the same two regimes, and $\gamma_\pm$ have the same values as for $\alpha_T$ \cite{eisler.unified}.

\begin{table}[htb]
	\centering
		\begin{tabular}{c|c|c|c}
		\hline
		Subj. & System & T/E & Refs. \\ \hline \hline
		\multirow{5}{*}{\begin{sideways}Networks\end{sideways}} & Random walk & T & \cite{barabasi.fluct, barabasi.separating,
 eisler.internal} \\ 
 		& Network models & T & \cite{tadic.loops, menezes.flux}\\ 
		& Highway network & T & \cite{barabasi.fluct, barabasi.separating}\\ 
		& World Wide Web & T & \cite{barabasi.fluct, barabasi.separating} \\
		& Internet & T & \cite{barabasi.fluct, barabasi.separating, duch.internet}\\ \hline
\multirow{2}{*}{\begin{sideways}Phy.\end{sideways}} & Heavy ion collisions & E & \cite{botet.prl,botet.pre, botet.npb} \\ 
		& Cosmic rays & E & \cite{uttley.flux, vaughan.flux} \\ 
\hline
		\multirow{5}{*}{\begin{sideways}Soc./econ.\end{sideways}} & Stock market & T & \cite{eisler.non-universality, eisler.unified, eisler.sizematters,jiang.fluxes} \\ 
		& Stock market & E & this review \\ 
		& Business firm growth rates & E & \cite{stanley.firm,amaral.growth} \\ 
		& Email traffic & T & this review \\
		& Printing activity & T & this review \\ \hline
		\multirow{2}{*}{\begin{sideways}Cl.\end{sideways}} & River flow & T & \cite{janosi.danube,dahlstedt.river} \\
		& Precipitation & T & \cite{eisler.inprep} \\ \hline
		\multirow{8}{*}{\begin{sideways}Ecology/pop. dyn.\end{sideways}} & Forest reproductive rates & T & \cite{ballantyne.scaling, ballantyne.correls} \\
		& Satake-Iwasa forest model & T & \cite{ballantyne.model}\\ 
		& Crop yield & T & \cite{fairfield.smith} \\
		& Animal populations & T, E & \cite{taylor,anderson.variability,conserv,maurer.taper} \\
		& Diffusion Limited population & E & \cite{kendal.dla} \\		
		& Population growth & T & \cite{keitt.pop, keitt.scaling} \\
		& Exponential dispersion models & E & \cite{kendal.ecological,kendal.blood,kendal.tumor} \\
		& Interacting population model & T & \cite{kilpatrick.ives} \\ \hline
		\multirow{9}{*}{\begin{sideways}Life sciences \end{sideways}} & Cell numbers & E & \cite{cells} \\
		& Protein expression & T & \cite{bareven.protein} \\
		& Gene expression & T & \cite{nacher.gene, tadic.gene} \\
		& Individual health & E & \cite{mitninski.ageing} \\
		& Tumor cells & E & \cite{kendal.tumor} \\
		& Human genome & E & \cite{genomevar, kendal.genomevar} \\
		& Blood flow & E & \cite{kendal.blood} \\
		& Oncology & E & \cite{kendal.tumor} \\
		& Epidemiology & T & \cite{measles,pathogen} \\
		\hline %
		\end{tabular}
	\caption{A list of some studies where fluctuation scaling/Taylor's law was directly applied or implied by a similar formalism. Groups were assigned by subject areas, Phy. = Physics, Cl. = Climatology. The column T/E shows the type of fluctuation scaling, T: temporal, E: ensemble.}
	\label{tab:examples}
\end{table}

\subsection{New empirical results}
\label{toc:2.4}
In this section we present previously unpublished results for fluctuation scaling.
{Note that temporal variances were estimated by the partition function of Detrended Fluctuation Analysis \cite{dfa}. This was necessary in order to (at least partly) remove the nonstationarity from the datasets. All results, including the values of $\alpha_\mathrm{T}$ agree qualitatively with those obtained from a direct calculation variance without detrending, but the accuracy of the estimation is improved.}

\subsubsection{Stock market (ensemble averaging)}
\label{toc:2.4.1}
Fluctuation scaling in stock market data has just been discussed, but those earlier results pertained temporal FS, whereas here we will present some new findings on ensemble FS in the same dataset.

We again consider the (daily) trading activity of stocks. The size of companies is often measured by the total value of all their issued stocks, called the company's capitalization ($C$). We take a fixed time period, the day 03/01/2000 (the results are similar for other days). Then we group the stocks according their capitalization into $35$ logarithmic bins. Finally, we calculate the mean $\evl{f_C}$ and the standard deviation $\evl{\sigma_C}$ of the activity in every group. EFS is shown in Fig. \ref{fig:Gstockensemble2}. The fit gives $\alpha_\mathrm{E} \approx 0.8-0.9$, although with some deviations from scaling.

If the size of the animal population is extensive, it is justifiable to use the size of the area to parametrize the ensemble: its mean should be exactly proportional to the area. To use capitalization as a parametrization for company size is a different matter. While it is indeed strongly related to the mean trading activity, there is no one-to-one correspondence between the two \cite{eisler.sizematters}. This means that companies of the same capitalization can have different expected trading activities. Thus our ensemble averaging technique is only approximate.

To circumvent this problem one can apply the following trick. The trading activity of companies fluctuates strongly from day to day, but the expectation value of the distribution is rather stable over time. So let us now take an interval $t=1\dots T$ and calculate the time averages $\ev{f_i(t)}$ during this period for each stock. Then for every stock take the single value $f_i(t=1)$ only, and group the observations according to $\ev{f_i}$. This is equivalent to the measurement before, only instead of $C_i$, groups are formed with respect to $\ev{f_i}$ ($15$ logarithmic bins). Then the ensemble mean $\evl{f}$ and variance $\evl{\sigma}$ can be calculated in each group. The results from this technique are also indicated in Fig. \ref{fig:Gstockensemble2}, one can see a dramatic decrease in the noise level, while the value of the exponent is approximately preserved, $\alpha_\mathrm{E} = 0.89$.

It must be emphasized that we use information about the temporal average only for the grouping procedure. The measured $\alpha_\mathrm{E}$ is a true EFS exponent. Moreover, the value of $\alpha_\mathrm{E}$ that we find here is much larger than $\alpha_\mathrm{T}$ (cf. Section \ref{sec:stock}). Because of the intricate statistical properties of markets, the two exponents cannot be expected to coincide. \footnote{The same two averaging techniques (fixed time and an ensemble of stocks versus a fixed stock and multiple times of observation) were previously introduced for stock market price changes in Ref. \cite{lillo.variety}.}

This example shows that the success of ensemble fluctuation scaling crucially depends on the proper choice of the size parameter. In the cases where possible, physical size or area are good choices, because they are known to be extensive. Otherwise the above trick can be applied, but only if multiple observations are available for every node and the system is close to stationary.

\addfig{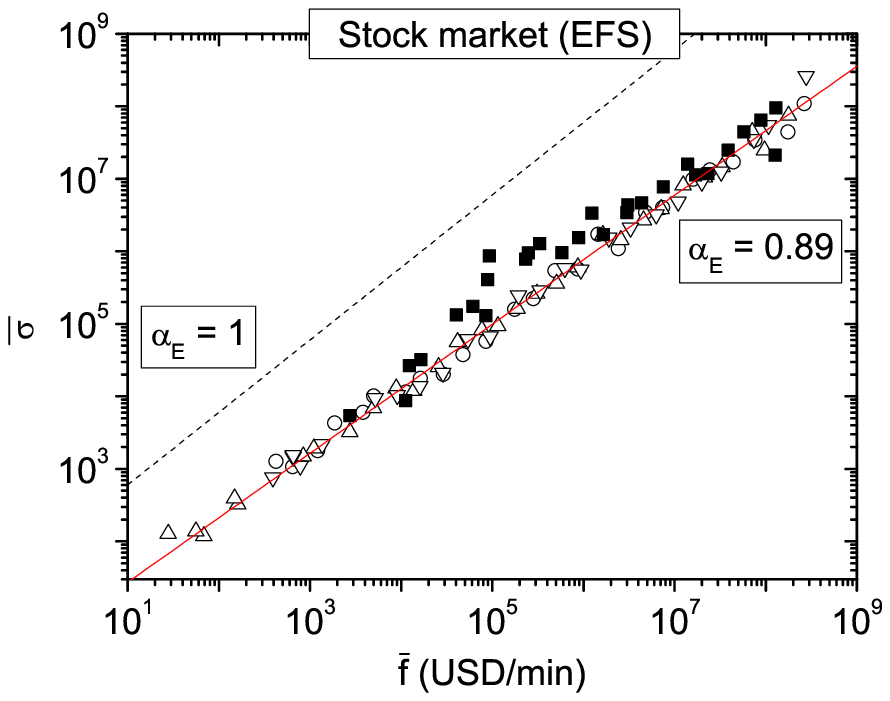}{{Fluctuation scaling considered as an ensemble average over NYSE stocks for the trading activity. Calculations were done both by considering the capitalization $C_i$ (03/01/2000: $\blacksquare$), and the monthly average trading activity $\ev{f_i}$ (03/01/2000: $\Circle$, 01/02/2000: $\bigtriangleup$, 01/03/2000: $\bigtriangledown$) as surrogates for size. One can see that using $\ev{f_i}$ for the formation of groups largely reduces fluctuations, but the exponent remains similar.}}

\subsubsection{Human dynamics}
\label{toc:2.4.2}
\label{sec:human}

The analysis of the records of human dynamics has recently seen growing interest \cite{barabasi.human, barabasi.einstein, vazquez.human}. Here we discuss two large technological databases of human activity:
\begin{enumerate}[(i)]
\item Emails from the employees of the company Enron during the year $2000$. We used a filtered variant of the original dataset posted by the Federal Energy Regulatory Commission \cite{enron.data}. We defined $f_i^{\Delta t}(t)$ as the number of emails sent by the person $i$ during the interval $[t, t+\Delta t)$.
\item Data on the printing activity of the largest printer at the Department of Computing at Imperial College London \cite{print.data}. The files include the complete year $2003$, we removed weekends, official holidays and closure times of the computer laboratory (23:00-7:00). We included $987$ users who submitted at least $3$ documents during our analysis, except the single largest user who appeared to have different statistical properties from the rest. $f_i(t)$ is defined as the number of documents submitted to print by user $i$ in the time interval $[t, t+\Delta t)$. Further details on the dataset can be found in Ref. \cite{paczuski.printing}.
\end{enumerate}
Note that multiple copies of the same email/document sent/submitted simultaneously were counted as one.

We are going to present these two datasets side by side, because they have strong similarities. Both show TFS for window sizes $\Delta t = 5\dots 2.8 \times 10^4$ sec. The exponent varies between $\alpha_\mathrm{T} = 0.52 - 0.72$ (email) and $\alpha_\mathrm{T} = 0.57 - 0.83$ (printing). Fig. \ref{fig:Genronalpha} shows the fits for time window sizes $\Delta t = 10, 10000$ sec. We find scaling over about $2.5$ orders of magnitude, and the exponent depends on $\Delta t$, as shown in Fig. \ref{fig:Genrondt}. Despite the level of the noise in the data, the dependence appears to be monotonically increasing, with two regimes separated by a crossover near $\Delta t \sim 4000$ sec (email) and $\Delta t \sim 1000$ sec (printing). The dependence is close to logarithmic, with the same form as for stock markets (the index $-$ corresponds to the regime below, $+$ above the crossover):
$$\alpha_{T,\pm}(\Delta t) = \alpha^*_{T,\pm} + \gamma_\pm \log \Delta t,$$ with $\gamma_-^\mathrm{email} \approx 0.04$, $\gamma_+^\mathrm{email} \approx 0.13$, $\gamma_-^\mathrm{print} \approx 0.09$, and $\gamma_+^\mathrm{print} \approx 0.02$. Also for the Hurst exponents $$H_{i;\pm} = H^*_\pm + \gamma_\pm \log \ev{f_i}.$$ The coefficients are $\gamma_-^\mathrm{email} \approx 0.04$, $\gamma_+^\mathrm{email} \approx 0.11$, $\gamma_-^\mathrm{print} \approx 0.07$, and $\gamma_+^\mathrm{print} \approx 0.01$. The $\sigma_i(\Delta t)$ scaling plots are shown in Fig. \ref{fig:Genronavg}, and the Hurst exponents' dependence on $\ev{f}$ in Fig. \ref{fig:Genronhurst}.

Finally, notice that for email data $\alpha_\mathrm{T}(\Delta t)$ tends to $1/2$ with decreasing window size, and the logarithmic tendency appears to saturate. On the other hand, for printing data the logarithmic tendency is markedly present even for short times. By an extrapolation from the trend one expects that $\alpha_\mathrm{T}(1\ \mathrm{sec}) \approx 0.51$. Section \ref{sec:alpha1/2} offers an explanation why for very short times one expects $\alpha_\mathrm{T} = 1/2$ in these datasets.

\addtwofigs{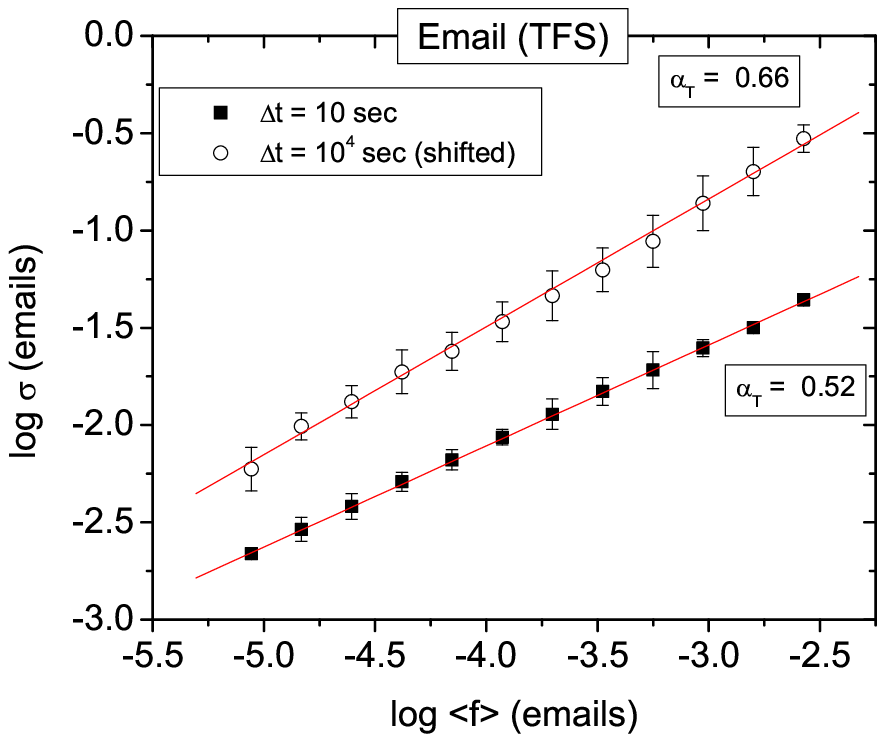}{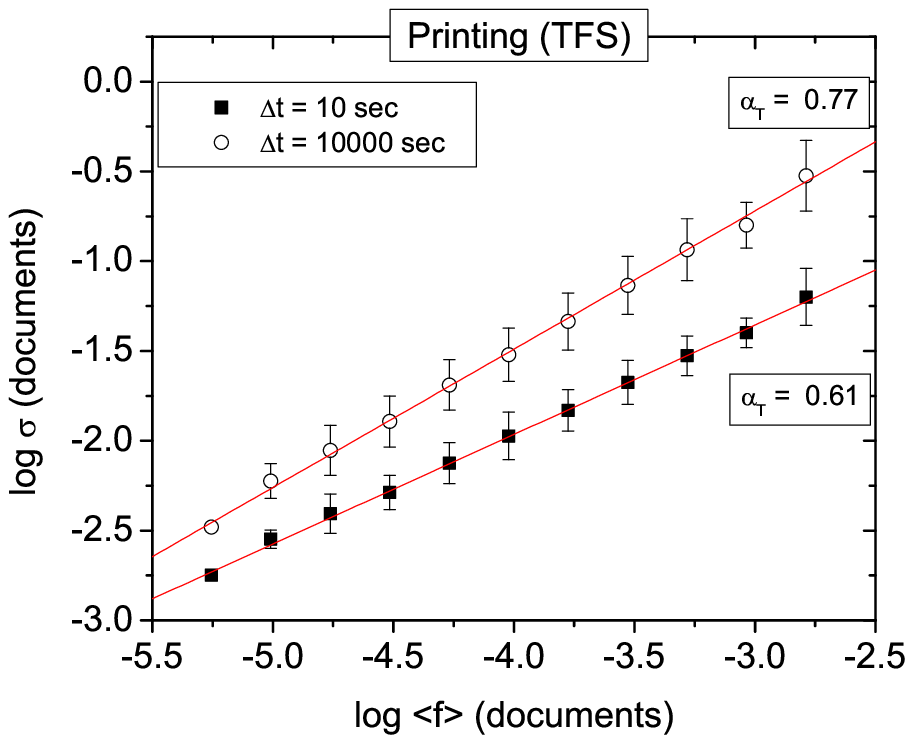}{Fluctuation scaling for time resolutions of $\Delta t = 10$ sec and $\Delta t = 10000$ sec. Points were logarithmically binned and $\log \sigma$ was averaged for better visibility, the error bars represent the standard deviations inside the bins. (left) Results for the number of sent emails. (right) Results for number of printed documents. }

\addtwofigs{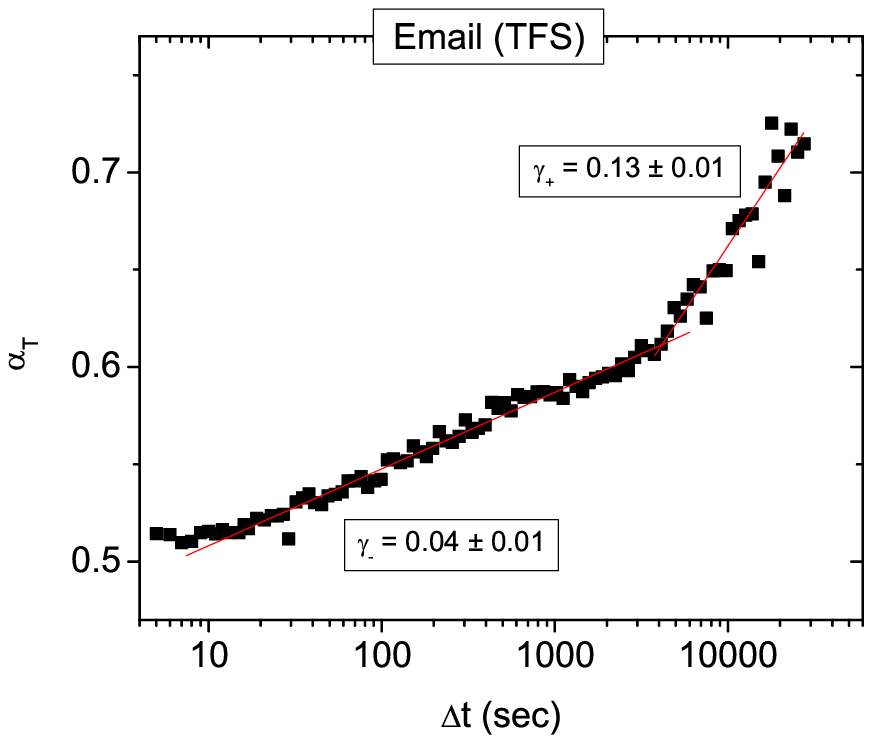}{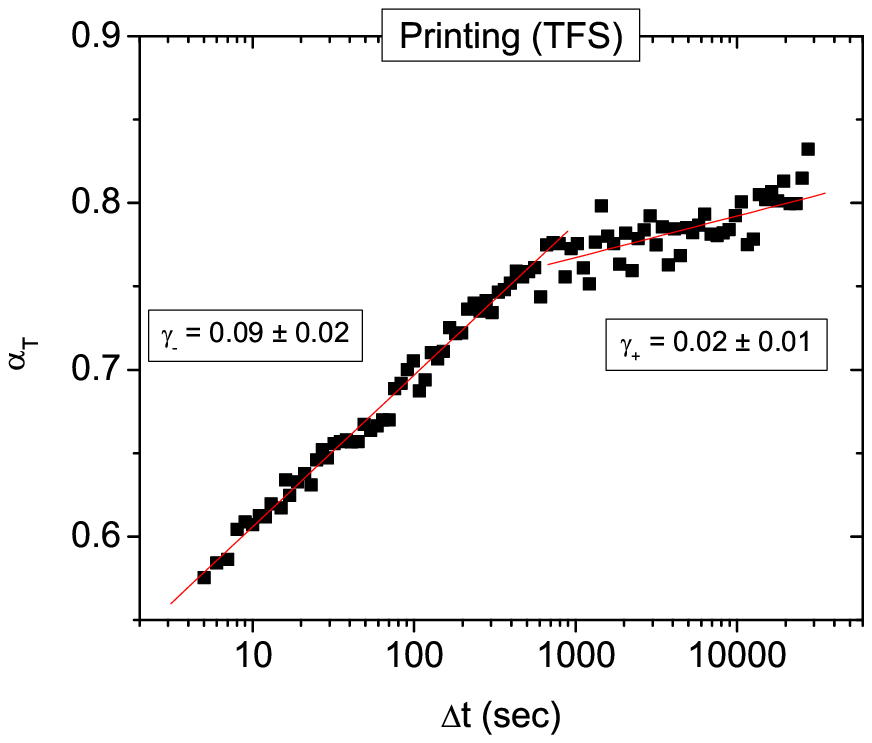}{(left) The dependence of the FS exponent in the Enron email database on the size of the time window $\Delta t$. The dependence is logarithmic in two regimes, with the coefficients $\gamma_- \approx 0.04$ for $\Delta t < 4000$ sec and $\gamma_+ \approx 0.13$ for $\Delta t > 4000$ sec. (right) The dependence of the FS exponent in printing data on the size of the time window $\Delta t$. The dependence is logarithmic in two regimes, with the coefficients $\gamma_- \approx 0.09$ for $\Delta t < 300$ sec and $\gamma_+ \approx 0.02$ for $\Delta t > 1000$ sec.}

\addtwofigs{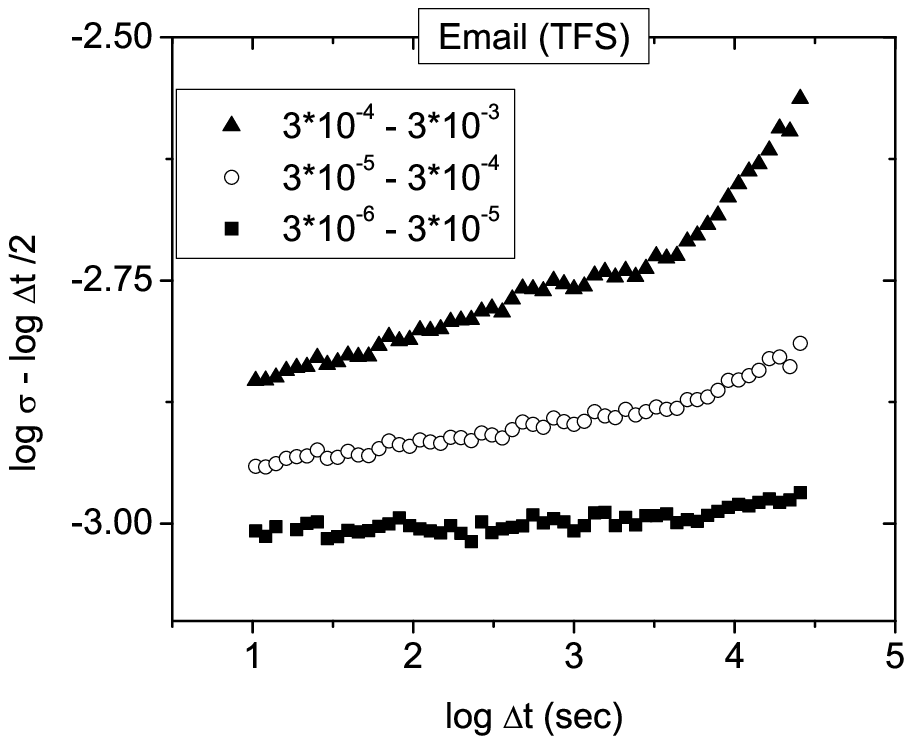}{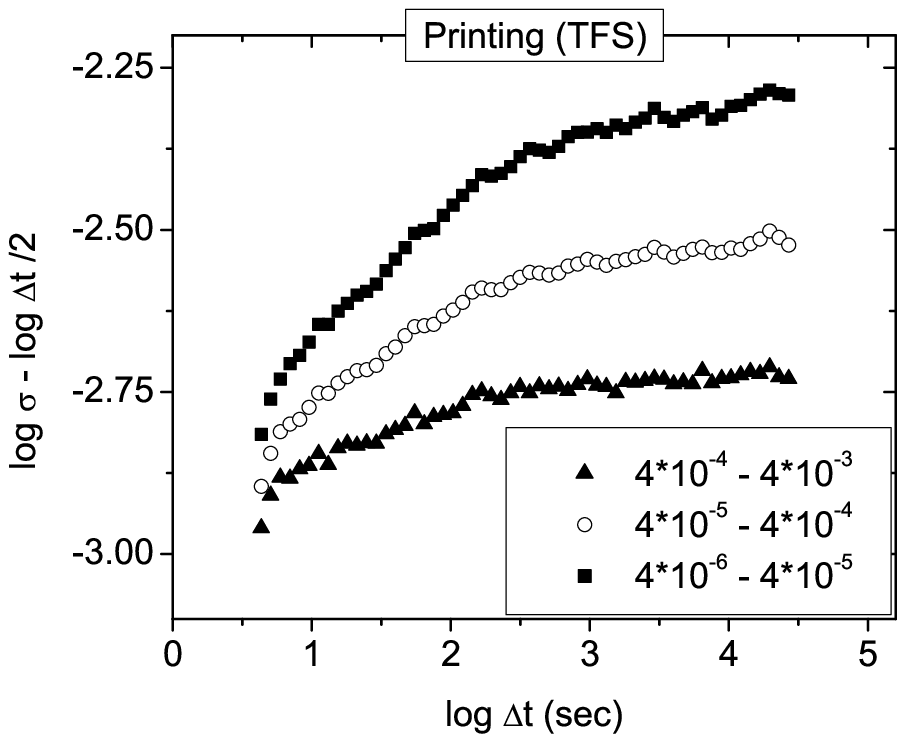}{Scaling plots of $\log \sigma - 1/2\times\log \Delta t$ versus $\log \Delta t$ generated by Detrended Fluctuation Analysis. Users were grouped by their average activity into three groups ($\ev{f}$ increasing from bottom to top, see plot for ranges) and the curves were averaged within groups. A horizontal line would correspond to complete the absence of correlations, and the slopes of the linear regimes are $H-0.5$, where $H$ are the typical Hurst exponents of groups. (left) Results for Enron email data. For shorter time windows $\Delta t < 4000$ sec, correlations are weak, and their strength increases slowly with greater $\ev{f}$. Then after a crossover regime, for $\Delta t > 4000$ sec correlations become stronger, with larger difference between the three groups. (right) Results for printing data. For shorter time windows $\Delta t < 300$ sec, positive correlations exist, and their strength increases with greater $\ev{f}$. Then after a crossover regime, for $\Delta t > 1000$ sec correlations become very weak for all three groups.}

\addtwofigs{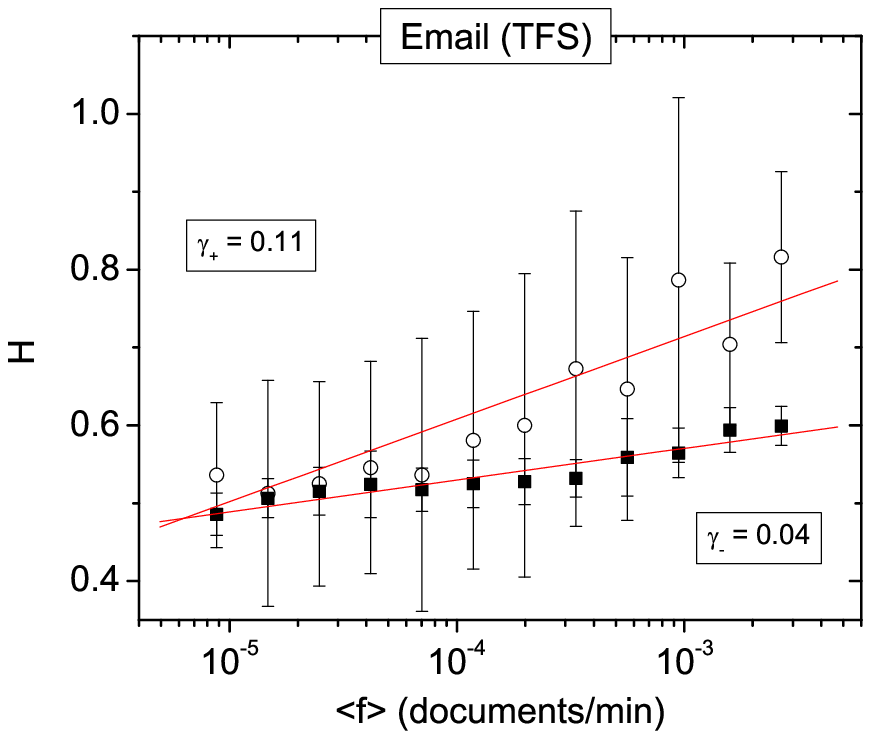}{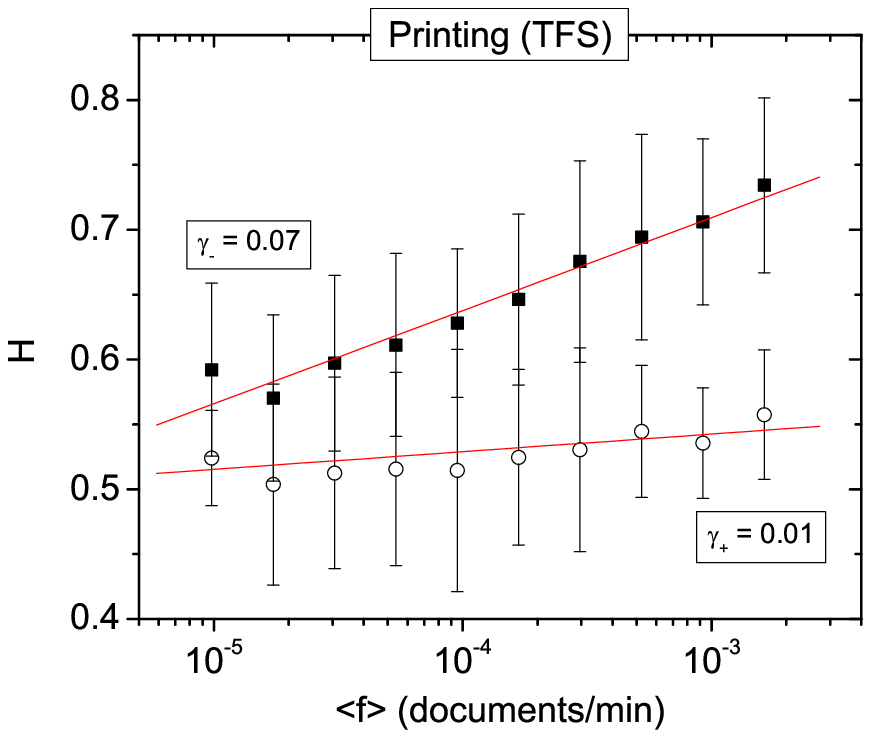}{The dependence of the Hurst exponent of $f$ on $\ev{f}$. Points were logarithmically binned and $H$ was averaged for better visibility, the error bars represent the standard deviations inside the bins. (left) Results for Enron email data. When $\Delta t < 4000$ sec ($\blacksquare$), depend weakly on $\ev{f}$, the coefficient of the logarithmic trend is $\gamma_- =0.04$, correlations are weak, $H_{i;-}\approx 0.5-0.55$. For time windows $\Delta t > 4000$ sec ($\Circle$), correlations become stronger, $\gamma_+ = 0.11$, but there are strong fluctuations around the logarithmic trend. (right) Results for printing data. When $\Delta t < 300$ sec ($\blacksquare$), the Hurst exponents depend logarithmically on $\ev{f}$ with a coefficient $\gamma_- =0.07$. For time windows $\Delta t > 10^3$ sec ($\Circle$), correlations are nearly absent, the respective Hurst exponents are $H_{i;+}\approx 0.5-0.55$, and the logarithmic dependence is weak, with $\gamma_+ = 0.01$.}

\subsubsection{Precipitation}
\label{sec:precipitation}
\label{toc:2.4.3}

In this section we present a study \cite{eisler.inprep} of the weekly
precipitation records of $22928$ weather stations worldwide. The
dataset was obtained from the Global Daily Climatology Network (GDCN) \cite{gdcn.data}. For one station, typically $40$ years of data is available between
$1950$ and $1990$. TFS is found with $\alpha_\mathrm{T} = 0.77$, see Fig.
\ref{fig:Lprecipitation}. However, there are also some significant
deviations. These can be interpreted based on geographical information: For every station, the geographical latitude ($l_i$, measured in degrees) and
the height $h_i$ measured from sea level was known, and the multiple
regression
\begin{equation}
   \log \sigma_i = C + \alpha_\mathrm{T} \log \ev{f_i} + C_l | l_i | + C_h h_i +
\epsilon_i
   \label{eq:precipreg}
\end{equation}
with an error term $\epsilon_i$ yields the results $C = 0.896 \pm
0.002$, $\alpha_\mathrm{T} = 0.732 \pm 0.002$, $C_l = -(8.79 \pm 0.05) \times
10^{-3}$ and $C_h = (-6 \pm 1 )\times 10^{-6}$. All values are
significantly different from zero at the $99.98\%$ confidence level.
For the single parameter fit of TFS, $R^2 =
0.73$, while for the multiple regression \eqref{eq:precipreg} one finds $R^2 =
0.90$ which is a substantial improvement due to the inclusion of geogrephical
latitude\footnote{Climatic fluctuations are well known to strongly
affect ecological fluctuations \cite{sheep, noisy.clockwork}. It is an
interesting fact that the variability of animal populations can also
depend on latitudinal position (see Ref. \cite{mcardle.variation} and
refs. therein).} and, in smaller part, to the height above sea level.
The remaining error term, although we did not find any appropriate
explanatory variable, is still not unsystematic. By plotting
$\epsilon_i$ on a map (see Fig. \ref{fig:Gprecipitationmap}) one
finds a strong geographical clustering with $\epsilon_i>0$ typically,
but not exclusively, in continental areas. This systematic tendency
suggests that a well-defined origin might exist for such corrections.

As for the value of $\alpha_\mathrm{T}$, its origin will be analyzed in a later
study \cite{eisler.inprep}. From preliminary studies it appears that
it is not strongly dependent on the choice of the time scale $\Delta
t$, and it is always significantly different from both $1/2$ and $1$.

\addfig{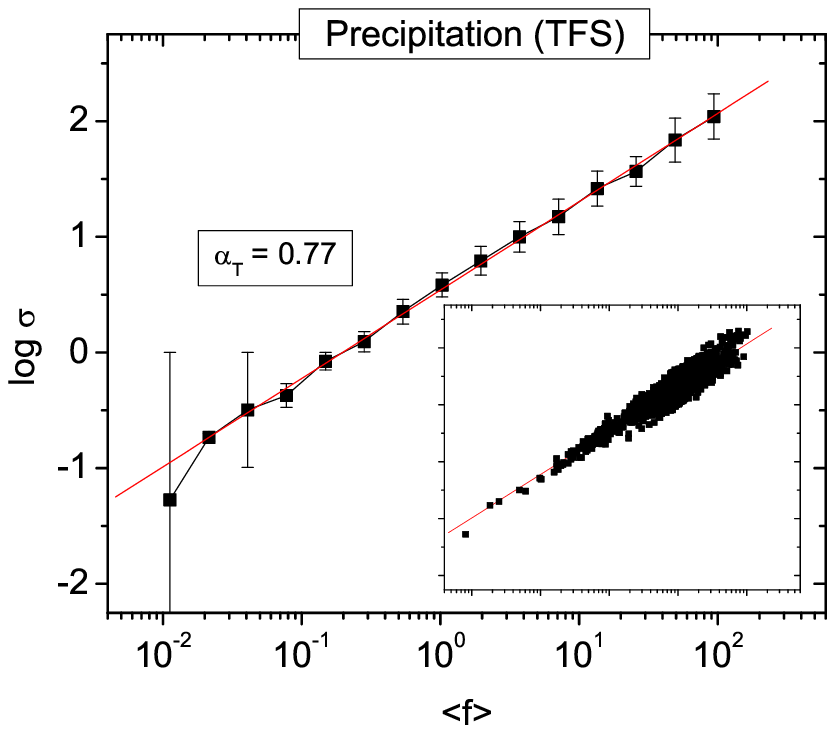}{Fluctuation scaling for the weekly precipitation of
weather stations. The fitted exponent is $\alpha_\mathrm{T} = 0.77$. Points were
logarithmically binned and $\log \sigma$ was averaged for better
visibility, the error bars represent the standard deviations inside
the bins. The inset shows the same plot and the same axis range, but
without binning. One can see that there is a high number of outliers,
due to other, $\ev{f}$-independent corrections to FS.}

\addwidefig{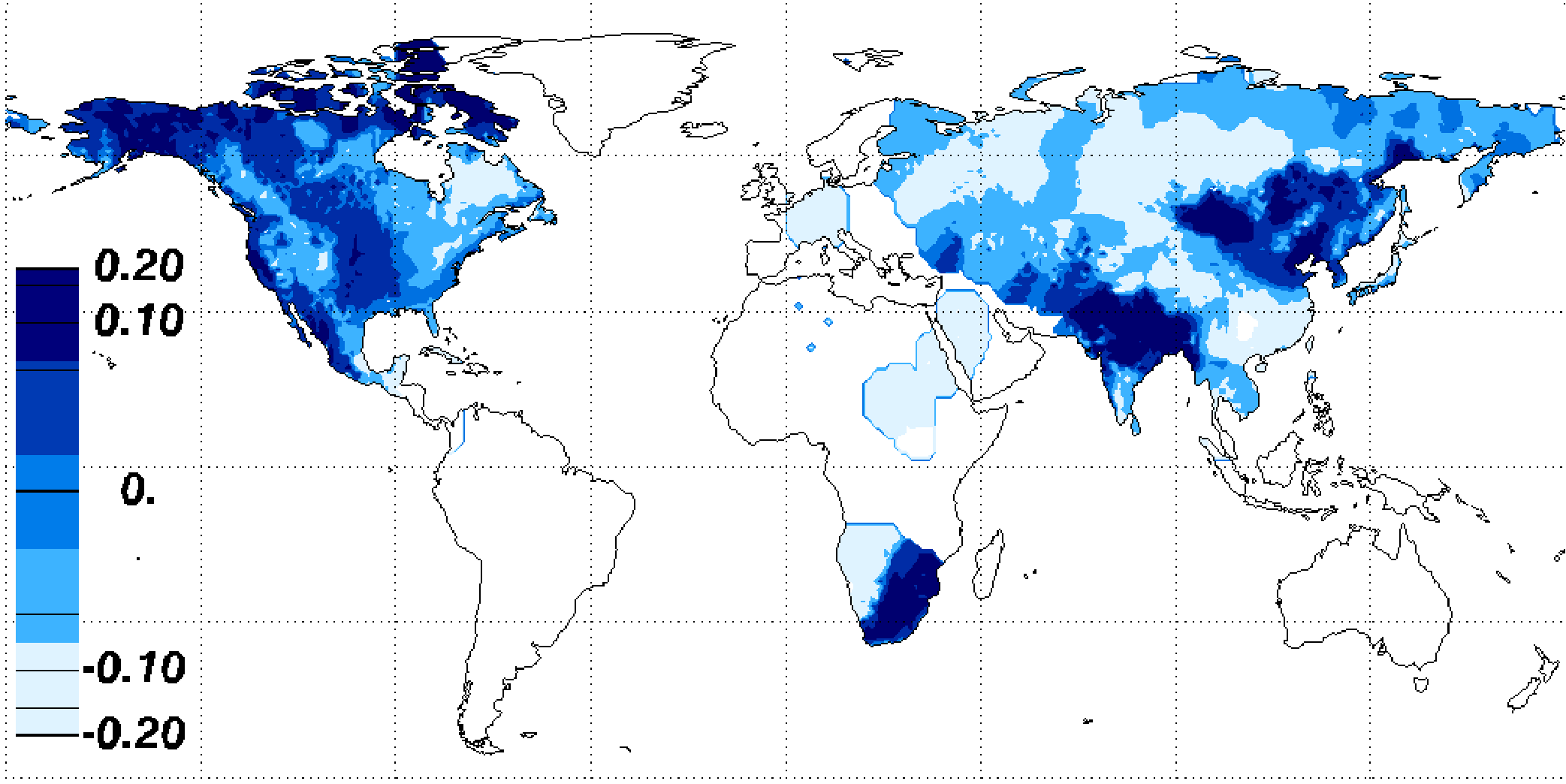}{The error terms $\epsilon_i$ of the
multiple regression \eqref{eq:precipreg}, which represent the
residuals of $\log \sigma$ after correcting for average precipitation,
geographical latitude and height above sea level. $\epsilon_i$
accounts for $10\%$ of the variance in $\log \sigma_i$, and it is
strongly clustered geographically suggesting the existence of a
well-defined underlying mechanism.}

\subsection{Corrections to fluctuation scaling}
\label{sec:corr}
\label{toc:2.5}

Similarly to precipitation data, there are significant corrections to TFS also in stock markets. These are related to differences between market sectors. In Fig. \ref{fig:Lstockcorrections} we plot $\sigma_i/\ev{f_i}^{\alpha_\mathrm{T}}$ versus $\ev{f_i}$ to characterize corrections to fluctuation scaling. By highlighting the alignment of three industrial sectors one can see that they form clusters, so the deviations are -- to some degree -- systematic.

In most real systems fluctuation scaling is found rather as a general tendency than an exact law. The scaling plots have some broadening, which can have several origins. A possible origin of poor fits can be the presence of crossovers and breaks in the scaling plots \cite{ballantyne.correls, anderson.variability, keeling.simplestoch, barabasi.fluct}, although these can only be seen clearly in very few studies \cite{botet.prl, janosi.danube}. In other cases the deviations are attributed to the quality of data and short sampling intervals \cite{smallsampletaylor}. Still, these corrections can be large and systematic. In Section \ref{sec:precipitation} we showed that for precipitation fluctuations geographical location and height plays a role. Similarly, for the stock market the market sector matters. These effects could be uncovered, because of the availability of these independent quantities for each station/stock. In some models they can even be calculated analytically, see Section \ref{sec:rwmanal}. 

The fact that scaling is mostly very well preserved suggests that the investigated complex systems have a robust dynamics characterized by a value of $\alpha$, and that the role of the corrections is not substantial in the formation of fluctuations. There is wide consensus that the exponents are meaningful, and not significantly distorted by the non-scaling corrections.

\addverywidefig{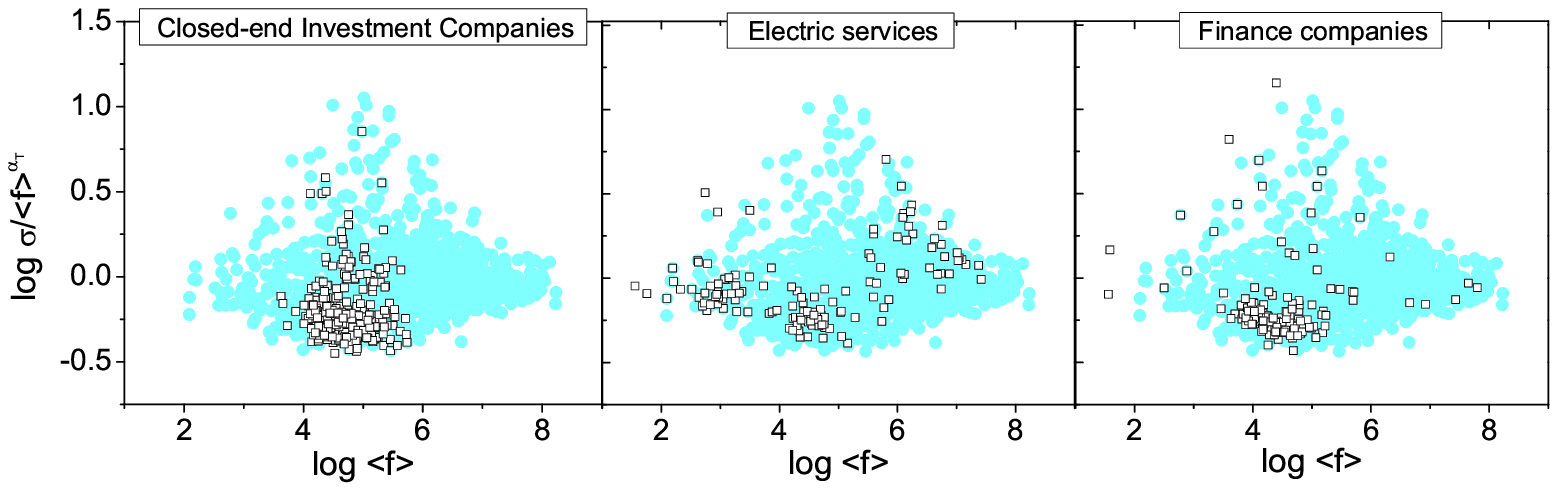}{{$\log[\sigma_i/\ev{f_i}^{\alpha_\mathrm{T}}]$ plotted versus $\log \ev{f_i}$ for the $10$-second resolution traded value data of stocks. The data points were not binned or altered in any way, which makes visible the deviations from the original scaling law, which would correspond to a horizontal line. The lighter points indicate all stocks, while boxes ($\Box$) highlight the distribution of points for the three indicated economic sectors. The great degree of clustering both horizontally and vertically. Clustering along the $\log \ev{f}$ axis only suggests that the sector has some typical trading activity. Systematic corrections to fluctuation scaling are indicated by clustering along the $\log \sigma/\ev{f}^{\alpha_\mathrm{T}}$ axis. If a sector is clustered in the lower/higher half of the dataset, it means that its trading activity has typically lower/higher fluctuations than the market average. The presence of such sector dependent clustering suggests that the corrections to fluctuation scaling are not purely random.}}

\subsection{Summary of observations}
\label{toc:2.6}

In sum, fluctuation scaling appears to be a surprisingly general concept that can be recognized in virtually any discipline where the proper data are available. The fluctuations of positive additive quantities appear to have the structure
$$\mathrm{fluctuations} = \mathrm{const.}\times \mathrm{average}^\alpha \times (1 + \mathrm{corrections}).$$
There is immense literature on the origins of fluctuations in various systems, ranging from gene networks \cite{paulsson.summing} through complexity \cite{barabasi.separating} to animal populations \cite{noisy.clockwork, engen.demographic, saether, sheep}. The common point of all these works is that fluctuations originate from two factors: internal and external. Naturally, the dynamics, the structure, and the interaction of the nodes vary from case to case. We expect that, e.g., the stock market trading activity and the reproduction of trees is very different. The discovery of FS as a common pattern can be a good start to point out further analogies and to build a broader picture.

\section{A general formalism}
\label{sec:gen}
\label{toc:3}

In the following we will focus on the \emph{temporal variant} of FS. In many cases the results continue to apply by simply dropping the time index $t$ and averaging over an ensemble of systems.

The previous section reviewed ample evidence that FS emerges in a very broad range of areas. Here we attempt to describe many of these by the same formalism. In the following we will assume that the systems are stationary. When considering a node $i$, its activity $f_i$ will always be decomposed as a sum. In some cases this means summation over the node's internal constituents, as for forests where reproductive activity was the total of that for all trees. In other cases the nodes themselves are simple, and the signal is the sum of events at the nodes, like the passing of cars at counting stations. If in the time interval $[t, t+\Delta t)$ there are $N_i^{\Delta t}(t)$ such constituents/events, and the activity of the $n$'th contributes $V_{i,n}^{\Delta t}(t)$ to the total activity $f_i$, then
\begin{equation}
	f_i^{\Delta t}(t) = \sum_{n=1}^{N_i^{\Delta t}(t)}V^{\Delta t}_{i,n}(t).
	\tag{\ref{eq:sum}}
\end{equation}
These $V$'s do not necessarily have to be independent \cite{ballantyne.correls, eisler.unified}, but we will assume that their (unconditional) distribution does not depend on $n$. We will omit the index $\Delta t$, where appropriate.

\subsection{The components of fluctuations}
\label{sec:components}
\label{toc:3.1}

Throughout this section we will analyze the fluctuations of quantities of the form \eqref{eq:sum}, as measured by the standard deviation/variance. Thus, it is important that there exists a simple analytical expression \cite{eisler.sizematters2} for $$\sigma^2_i = \ev{\left [ f_i(t) - \ev{f_i(t)}\right]^2}.$$ Appendix \ref{app:components} gives a proof that
\begin{equation}
\sigma^2_i = \Sigma^2_{Vi} \ev{N_i^{2H_{Vi}}} + \Sigma^2_{Ni} \ev{V_i}^2,
\label{eq:comp}
\end{equation}
where $\ev{V_i}$ and $\Sigma^2_{Vi}$ are the mean and the variance of $V_{i,n}$. Similarly, $\ev{N_i}$ and $\Sigma^2_{Ni}$ are the mean and variance of $N_i$. We also introduced the Hurst exponent $H_{Vi}$ of the constituents, which is defined as
\begin{equation}
\Sigma^2_{Vi}(N)=\ev{\left [\sum_{n=1}^N V_{i,n} -\ev{\sum_{n=1}^N V_{i,n}}\right ]^2} \propto N^{2H_{Vi}}
\label{eq:tickhurst}
\end{equation}
for any $t$ and $\Delta t$. If for a fixed $i$ and $t$ all $V_{i,n}(t)$ are uncorrelated, then $H_V = 1/2$, while if they are long range correlated $H_V > 1/2$ \cite{hurst, dfa}. \footnote{The Hurst exponent is only related to correlations in this simple fashion because the distribution of $V_{i,n}(t)$ does not depend on $n$ \cite{scalas.contt2}.}

\subsection{"Universality" classes}
\label{toc:3.2}
\label{sec:univcl}

In this formalism it is relatively easy to show that there exist two important classes of systems, one with $\alpha = 1/2$ and one with $\alpha = 1$. The existence of such classes was pointed out, for example, by Anderson et al. \cite{anderson.variability} and later by Menezes and Barab\'asi \cite{barabasi.fluct}. 

These are \emph{not universality classes} but rather simple limiting cases, and $\alpha$ is not a universal exponent in the usual sense of statistical physics \cite{reichl}. Many empirical systems do not belong to either class, and both $\alpha=1/2$ and $\alpha=1$ can arise from several types of dynamics. In order to make FS a truly useful tool in the analysis of empirical data, one needs a classification scheme for how different types of dynamics can be mapped onto $\alpha$. Our current understanding of such classification is outlined in this section.

\subsubsection{The case $\alpha = 1/2$}
\label{sec:alpha1/2}
\label{toc:3.2.1}

We will now present two scenarios that can give rise to $\alpha = 1/2$. The arguments will be given in the language of time averages, but they can be generalized to ensemble averages in a straightforward way.

\begin{enumerate}
\item[1)] Let us assume that every node $i$ consists of a fixed number $N_i(t)=N_i$ of constituents, each with a signal $V_{i,n}(t)$ which is i.i.d. for all $i$, $n$ and $t$, with the same mean $\ev{V}$ and variance $\Sigma^2_V$. From Eq. \eqref{eq:comp} it is trivial that here $\sigma^2_i = N_i\Sigma^2_V.$ Because of the linearity of the mean, $\ev{f_i}=N_i\ev{V},$ so $$\sigma^2_i = \frac{\Sigma^2_V}{\ev{V}}\ev{f_i},$$ and hence $\alpha = 1/2$. Of course in this simple case one can say more, because the \emph{central limit theorem} \cite{feller} is applicable:
\begin{equation}
	\frac{\sum_{n=1}^{N_i}V_{i,n}(t)-N_i\ev{V}}{\sqrt{N_i}\Sigma_V}\rightarrow\mathcal G_i(t),
\end{equation}
where $\mathcal G_i(t)$ are i.i.d. standard Gaussians and "$\rightarrow$" means convergence in distribution for $N_i\rightarrow\infty$. 

Exactly the same equation can be rewritten to more resemble fluctuation scaling:
\begin{equation}
	\frac{f_i(t)-\ev{f_i}}{K\ev{f_i}^\alpha} \rightarrow \mathcal G_i(t).
\end{equation}
The power $\alpha=1/2$ is exactly the power in FS, and $K=\Sigma_V\ev{V}^{-1/2}$. The conceptual difference is only that since we know that $\ev{f_i}=N_i\ev{V}$, we can use $\ev{f_i}\ev{V}^{-1}$ as a surrogate variable for $N_i$. This is very useful, when we only have the time series of $f_i(t)$ available but not $N$, since the limit can be switched to $\ev{f}\rightarrow\infty$ (cf. Section \ref{sec:limit}).

\item[1')] Let us consider an example for scenario 1): a system, where $V_{i,n}(t)$ can only be $1$ with probability $p$ and $0$ with probability $1-p$. Scenario 1) still applies because $V$'s are i.i.d., so $\alpha = 1/2$. This binary distribution can be instructive, as one can think of $V_{i,n}(t)$ as \emph{independent indicator variables}. For example let us take a volume $S$ of ideal gas within a large container. Let the whole system contain $N$ gas atoms, and $V_{i,n} = 1$ if the $n$'th atom is in the container, while $V_{i,n} = 0$ if it is not. The ideal gas is homogeneous and the atoms are independent, every atom having a probability $p\propto S$ of being in the small container. From here, one can apply the above argument to show that for various containers of different sizes $$\sigma_S \propto \ev{f_S}^{1/2}.$$
These are the well-known square-root type fluctuations of equilibrium statistical physics (see, e.g., Section XII of Ref. \cite{landau5}). Similar arguments were suggested for the number of animals in an area: If the motion of individuals were independent (gas-like), then their spatial density fluctuations should follow $\alpha = 1/2$ \cite{taylor.woiwod}.

{\item[1'')] The example of the ideal gas can be given in the language of ensemble averages as well. Simply we take a large number of containers of the same size $S$ and calculate the mean $\overline{f_S}$ and standard deviation $\overline{\sigma_S}$ of atom counts between these containers. Then we vary the container size, and we recover an analogous relationship:
$$\overline{\sigma_S} \propto \overline{f_S}^{1/2}.$$
Of course, this was expected, because the system is ergodic, so temporal and ensemble averages are equal.}

\item[2)] For an even simpler mechanism let us recall the findings of Section \ref{sec:human}. We found that for very short times ($\Delta t \sim 1$ sec) the number of sent emails/printed documents follow TFS with $\alpha_\mathrm{T} = 1/2$. It is highly unlikely that someone will send several different emails/print several different documents in the same second (duplicates of the same email to multiple recipients were excluded). Thus $f_i(t) = 0$ or $1$, and so $f_i(t) = f_i^2(t)$. Remember that this is very different from the previous example, where $f_i(t)$ was allowed to have any value, and only $V_{i,n}(t)$'s were constrained to $0$ or $1$.

In the email/print data the number of events per second was very low, generally $\ev{f_i} < 4\times 10^{-3}$ sec$^{-1}$. The standard deviation is then
\begin{equation}
	\sigma_i^2 = \ev{f_i^2}-\ev{f_i}^2 = \ev{f_i}-\ev{f_i}^2 \approx \ev{f_i},
\end{equation}
so $\alpha = 1/2$. The same argument holds for the number of trades per second in the stock market \cite{eisler.unified}.

The meaning of this scenario 2) in summary: We are examining the system on such a short time scale that no two events happen in the same time window. Then, the FS exponent tells us nothing about the dynamics of the system, because $\alpha = 1/2$ is automatically true.
\end{enumerate}

\subsubsection{The case $\alpha = 1$}
\label{sec:alpha1}
\label{toc:3.2.2}

We will now present two scenarios that can give rise to the value $\alpha = 1$. While 1) is only valid for TFS, 2) can be readily generalized for EFS as well.

\begin{enumerate}[1)]
\item[1)] It was possible to obtain $\alpha = 1/2$ by sums of \emph{independent} $V$'s. In the other extreme case, if every node $i$ had a fixed number of \emph{identical and completely synchronized} constituents, i.e., $N_i(t)=N_i$ and $V_{i,n}(t)\equiv V_i(t)$ Eq. \eqref{eq:sum} simplifies to
$$f_i(t) = \sum_{n=1}^{N_i}V_{i,n}(t) = N_iV_i(t).$$
Then $\ev{f_i(t)}=N_i\ev{V_i(t)}$, and $\sigma_i = N_i\Sigma_{Vi}$.
\begin{equation}
	\sigma_i = \frac{\Sigma_{Vi}}{\ev{V_i}}\ev{f_i}\propto \ev{f_i}^\alpha,
\end{equation}
with $\alpha = 1$. The last proportionality only holds if the ratio ${\Sigma_{Vi}}/{\ev{V_i}}$ is the same for any $i$, for example when the distribution of $V_i$ is independent of $i$.\footnote{If the dependence is present but weak, then it may cause corrections to FS, but scaling should still hold approximately.}

\item[1')]{How is such an argument of any use? The study of Cho et al. \cite{cho.genome} reports experimental data on samples of yeast, in which cells were artifically prepared to have almost perfectly synchronized cell cycles. The measured signal $f_i(t)$ is the hourly expression level of various genes $i$ in a sample. If all cells of yeast contribute in the same way to the measured expression level, and they are synchronized, then the value $\alpha_\mathrm{T} = 1$ is simply an indicator of such a synchrony. Thus FS probably tells us nothing about the dynamics of gene transcription, and the exponent is simply due to the sample preparation.}

{Nacher et al. \cite{nacher.gene} propose a stochastic differential equation model that predicts the same exponent $\alpha_\mathrm{T} = 1$ for this dataset ($\alpha_\mathrm{T} = 1$ is confirmed by Zikovi\'c et al. \cite{tadic.gene}). They argue that self-affine temporal correlations are the origin of such a value. Section \ref{sec:gamma} will show that self-affine temporal correlations do not contribute to $\alpha_\mathrm{T}$ in this way. Instead, our above explanation is simpler, and it suggests that the dataset cannot be used in favor of any proposed model based on the value of $\alpha_\mathrm{T}$.}

\item[1'')] Real systems are often not closed, but subject to outside forces. In certain cases this driving can be so strong that it can overwhelm the internal dynamics. If the internal structure of the system becomes irrelevant, this must also have an effect on FS. There have been a number of studies discussing how fluctuations in complex systems are formed as the sum of internally generated and externally imposed factors \cite{paulsson.summing, noisy.clockwork, engen.demographic, saether, sheep}. Anderson et al. \cite{anderson.variability} and Menezes and Barab\'asi \cite{barabasi.fluct, barabasi.separating} suggested that $\alpha_\mathrm{T} = 1$ can arise when the external driving force imposes strong fluctuations in either $V_i(t)$ or $N_i(t)$ (cf. Ref. \cite{noisy.clockwork}).

When all $V_{i,n}(t)$ (the signals of every constituent at every node) become synchronized, then we are back at scenario 1): $\alpha = 1$, because $f_i(t)/\ev{f_i}=V(t)/\ev{V}$ which has a universal, $i$-independent distribution.

It is also possible that an external force $W(t)$ affects the number of constituents in the elements so strongly that the fluctuations of $N_i(t)$ become proportional only to this force. In this case $N_i(t) = A_i W(t)$, where $A_i$ are node-dependent constants. {One expects that generally $\ev{f_i(t)}=A_i\ev{W(t)}\ev{V_i}$, whereas
$$\sigma_i^2 = \Sigma^2_{Ni}\ev{V_i}^2+\Sigma^2_{Vi}\ev{N}=A_i^2\Sigma^2_W\ev{V_i}+\Sigma^2_{Vi}A_i\ev{W}.$$
If fluctuations in $W$ are so large that $\ev{W} \ll \Sigma_W^2$, then only the first term remains. After some algebraic steps
	$$\sigma_i^2 \approx \frac{\Sigma^2_W}{\ev{W}^2\ev{V_i}}\ev{f_i}^2 \propto \ev{f_i}^{2\alpha_\mathrm{T}},$$
with $\alpha_\mathrm{T} = 1$. The last proportionality is true if the distribution of $V_{i,n}$ does not depend strongly on $i$.}

\item[2)] $\alpha = 1$ can be a sign of a universal distribution of $f_i(t)/\ev{f_i}$, which only varies by a constant multiplicative factor throughout nodes. If this is true, then $f_i(t)$ can be decomposed into this factor $F_i$, and the universal random variable $V_i(t)$, which are identically distributed for all $i$. Naturally $\ev{f_i}=F_i\ev{V}$, and $\sigma^2_i=F_i^2\Sigma^2_V$, and $\sigma_i = \Sigma_V\ev{V}^{-1}\ev{f_i}$.
\end{enumerate}

\subsection{Other values of $\alpha$}
\label{toc:3.3}

It has been observed that many real systems obey FS with $\alpha$ values that significantly differ from both $1/2$ and $1$. In this section we summarize the current knowledge of general mechanisms that can give rise to intermediate values.

\subsubsection{The dependence of $\alpha$ on the time resolution $\Delta t$}
\label{sec:gamma}
\label{toc:3.3.1}

First of all, $\alpha$ can depend on the size of the time window used for its measurement. This phenomenological picture can be used to understand the results of Section \ref{sec:stock} for stock market trading, and Section \ref{sec:human} for human activity.

Let us assume, that the activity time series are long time correlated with Hurst exponents $H_i$ that are allowed to depend on the node $i$. The Hurst exponent of the time series $f_i^{\Delta t}(t)$ was previously defined as
\begin{equation*}
\sigma_i(\Delta t) = \ev{\left [ f_i^{\Delta t}(t) - \ev{f_i^{\Delta t}(t)}
\right ]^2}^{1/2} \propto \Delta t^{H_i}. \tag{\ref{eq:hurst}}
\end{equation*}
This definition is almost exactly the same as Eq. \eqref{eq:tickhurst} for $H_V$, the only difference being that now instead of the $N$ number of constituents we consider the time window size $\Delta t$ as the scaling variable. TFS deals with how the variance scales when one moves to stronger (larger $\ev{f}$) signals:
\begin{equation*}
	\sigma_i \propto \ev{f_i}^\alpha \tag{\ref{eq:taylori}}.
\end{equation*}
Eq. \eqref{eq:hurst} takes an alternative point of view and suggests that for a fixed signal, in the presence of long-range temporal correlations, the variance can grow anomalously also by changing the time window.

Following Ref. \cite{eisler.unified}, from Eqs. \eqref{eq:hurst} and \eqref{eq:taylori}, it is easy to see that the roles of $\ev{f_i}$ and $\Delta t$ are analogous. Since the left hand sides are the same, one can write a third proportionality between the right hand sides:
$$\Delta t^{H_i} \propto \ev{f_i}^{\alpha(\Delta t)}.$$
After taking logarithm on both sides, and differentiating by
$\partial^2/\partial(\log \Delta t)\partial(\log \ev{f_i})$, one finds that asymptotically
\begin{equation}
\frac{d H_i}{d (\log\ev{f_i})}\sim\frac{d \alpha(\Delta t)}
{d (\log\Delta t)}\sim \gamma.
\label{eq:gammagener}
\end{equation}
This means that both partial derivatives have the same constant value,
which we will denote by $\gamma$.

Eisler and Kert\'esz \cite{eisler.unified} outline three scenarios for this equality to hold:
\begin{enumerate}[(I)]
\item In systems, where $\gamma = 0$, the exponent
$\alpha$, is independent of window size, and the degree of temporal correlations ($H$)
is the same at all nodes.
\item When $\gamma > 0$, $\alpha(\Delta t)$ depends on $\Delta t$ logarithmically:
$\alpha(\Delta t) = \alpha^* + \gamma_1 \log \Delta t$. The Hurst exponent of the node also depends on $\ev{f}$ logarithmically with the same prefactor:
$H_i = H^* + \gamma\log\ev{f_i}$.
\item It is possible that Eq. \eqref{eq:gammagener} only holds piecewise, for certain ranges in $\Delta t$. Two regimes are then separated by a crossover between two distinct values $\gamma_\pm$, and nodes will have separate Hurst exponents $H_-(i)$ and $H_+(i)$ in the two regimes.
\end{enumerate}
Case (III) was shown for the stock market and human dynamics in Section \ref{sec:empiricaltime}.

\subsubsection{Impact inhomogeneity}
\label{sec:beta}
\label{toc:3.3.2}

Any value of $\alpha$ can easily arise without dependence on the time window. To better understand the reason how, consider three toy systems with the following elements.
\begin{enumerate}[(i)]
	\item Let us take a fair coin with $0$ written on one side and $1$ on the other, this will be our group $i=1$. Then take two such coins for group $i=2$, three for $i=3$, etc. In every time step we flip all coins in every group, and let $f_i$ equal the sum of the numbers we flipped in element $i$. Naturally $\ev{f_i} \propto i$ and, if all coins are independent, $\sigma_i \propto i^{1/2}$. Thus, for such a case $\alpha = 1/2$.
	\item Now let us take another fair coin with $0$ written on one side and $1$ on the other, this will be our element $i=1$. For $i=2$, we again take only one coin with sides $0$ and $2$. For any $i$, there will be a single coin with sides $0$ and $i$. Trivially $\ev{f_i}\propto i$, but also $\sigma_i \propto i$. So this time $\alpha = 1$.
	\item In our final example, let us mix the above two. For the $i$'th group there are $i$ coins, each having a side with $0$ and a side with $i$. Then $\ev{f_i}\propto i^2$, whereas $\sigma_i \propto i^{1/2}\times i$. We have just constructed a case for $\alpha = 3/4$.
\end{enumerate}

One can unify these examples by introducing \emph{impact inhomogeneity}. One can write the contribution (impact) of the constituents at a node $i$ as 
\begin{equation}
	V_{i,n}(t) = \ev{V_{i,n}}\cdot X_{i,n}(t),
	\label{eq:vinoisy}
\end{equation}
all $X_{n,i}(t)$ are i.i.d. with unit mean. We then allow $\ev{V_{i,n}}$ to depend on $\ev{N_i}$ as a power law between nodes \cite{eisler.internal, keitt.scaling}:
\begin{equation}
	\ev{V_{i,n}}\propto \ev{N_i}^\beta.
	\label{eq:beta}
\end{equation}
According to Eq. \eqref{eq:comp} fluctuations can be calculated as
\begin{eqnarray*}
	\sigma^2_i = \Sigma^2_{Vi} \ev{N_i} + \Sigma^2_{Ni} \ev{V_i}^2 = \\ \Sigma^2_X \ev{V_i}^2\ev{N_i} + \ev{N_i}\ev{V_i}^2 \propto \ev{f_i}^{2\alpha},
\end{eqnarray*}
where $\Sigma^2_X = \ev{X^2}-\ev{X}^2$, and
\begin{equation}
	\alpha = \frac{1}{2}\left(1+\frac{\beta}{\beta+1}\right),
	\label{eq:alphabeta}
\end{equation}
where we introduced the new parameter $\beta$.

As a quick check, the three toy models correspond to $\beta = 0$, $\alpha = 1/2$ (all coins $0$ or $1$); $\beta = 1$, $\alpha = 3/4$ (the coins value proportional to their number) and $\beta \rightarrow \infty$, $\alpha = 1$ (only one coin with growing value). There is always some $\beta \geq 0$ that allows us to reproduce a given value $\alpha \in [1/2, 1)$, whereas the range $\beta < 0$ covers all possibilities of $\alpha < 1/2$ and $\alpha > 1$.

\subsubsection{Examples of impact inhomogeneity}
\label{sec:tweedie}
\label{toc:3.3.3}

The ecology literature has documented \cite{gaston.lawton, gaston.range} that empirically there is a strong positive correlation between the typical size of subpopulations\footnote{This is often called local abundance.} ($\ev{V}$) and the number of subpopulations per unit area ($\ev{N}$) or the total population per unit area\footnote{regional abundance} ($\ev{f}$). The conjecture that these quantities might behave as powers of each other as in Eq. \eqref{eq:beta} was proposed by Keitt et al. \cite{keitt.scaling}, both across species and for individual subpopulations of the same species.

Kendal makes a similar suggestion, and shows that it generates non-trivial exponents in EFS for ecological populations \cite{kendal.ecological} and the heterogeneity of blood flow in organs \cite{kendal.blood}. In fact he does not point out the general mechanism, but instead refers to non-trivial EFS exponents as the property of a class of models, which entail impact inhomogeneity. Here we will omit most of the formalism; a proof that Kendal's approach has impact inhomogeneity can be found in Appendix \ref{app:tweedie}.

Let us take the case of animal populations as the example. Kendal proposes that EFS holds with an exponent $1/2<\alpha<1$ if the population of an area can be described by the so-called Tweedie exponential dispersion models \cite{penis}. These assume that (i) an area $i$ contains a Poisson distributed number of animal clusters ($N_i$), (ii) the size of individual clusters ($V_{i,n}$) is i.i.d. gamma distributed, (iii) and there is a power law relationship between the means of these two quantities. Of course, (iii) is the same as Eq. \eqref{eq:beta}, along with all of its consequences.

As for blood flow \cite{kendal.blood}, it is measured by the entrapment of radioactive microspheres in capillaries. In a fixed mass of tissue, the number of entrapment sites $N$ is assumed to be Poisson distributed, while the blood flow $V$ of the sites is taken as gamma distributed, along the same lines and with the same conclusions as above.

Finally, Section \ref{sec:stock} suggested impact inhomogeneity also as the origin of non-trivial FS exponents for the traded value on stock markets.

\subsubsection{Constituent correlations}
\label{sec:sync}
\label{toc:3.3.4}

There exists a further mechanism to produce any value $1/2\leq\alpha\leq1$, without considering the scaling property of impacts. The total output of node $i$ is given by the equation
\begin{equation}
	f_i(t) = \sum_{i=1}^{N_i}V_{i,n}(t). \tag{\ref{eq:sum}}
\end{equation}
We also fix $N_i$ as time independent. If we assume that the unconditional distribution of $V$'s is independent from $N_i$, and also from $n$, then one can denote the expectation value $\ev{V}=\ev{V_{i,n}(t)}$.

The central idea is the introduction of \emph{correlations between constituents, i.e., variables with different $n$}. Let us assume for simplicity that the elements are situated on a one-dimensional lattice, and their activity is long-range correlated in space, so that the correlation function decays as a power law,
\begin{equation}
	C(\Delta n) \propto \ev{V_{i,n}V_{i,n+\Delta n}}-\ev{V_{i,n}}^2 \propto \Delta n^{2H_V-2}.
	\label{eq:Cdn}
\end{equation}
$H_V$ is the same Hurst exponent, as defined in Eq. \eqref{eq:tickhurst}. Then, positively correlated patterns display $H_V>1/2$, for uncorrelated (or short range correlated) patterns $H_V=1/2$, and for anticorrelated (antipersistent) patterns $H_V<1/2$. \footnote{We need to assume that $V_{i,n}$ is stationary as a function of $n$. }

It follows from Eq. \eqref{eq:comp} that the fluctuation of the combined activity of all constituents is:
$$\sigma^2_i=\Sigma^2_{V}N_i^{2H_V} \propto \ev{f_i}^\alpha,$$
where
\begin{equation}
	\alpha = H_V.
	\label{eq:popalpha}
\end{equation}
This idea was (to our knowledge) first presented by West \cite{west.comments}, and demonstrated on surrogate data sets, but it was not applied directly to any new problem. The role of spatial correlations in the formation of FS in the context of ecology was also suggested by Colman et al. \cite{colman.regulated} and Ballantyne and Kerkhoff \cite{ballantyne.correls} more recently. The idea is confirmed by simulations, see Section \ref{sec:binary}.

\section{Models}
\label{sec:mod}
\label{toc:4}

In this section, we will discuss some models that can be used to understand basic facts about fluctuation scaling, how it arises and what its limitations are. 

\subsection{Random walks on complex networks}
\label{toc:4.1}
\label{sec:rw}

\subsubsection{The model}
\label{toc:4.1.1}
It was proposed by Menezes and Barab\'asi \cite{barabasi.fluct} that random walks can generate TFS in the following way. Let us take a scale free Barab\'asi-Albert network\footnote{The particular topology is irrelevant from the point of view of $\alpha_\mathrm{T}$. The network only has to be connected and the nodes should have a wide range of degrees.} of $M$ nodes \cite{barabasi.rmp}. We distribute $W$ independent random walkers (tokens) randomly to the nodes. Then, in every time step these jump from their current node to one of their neighbors randomly. The process is repeated for $s=1\dots s_\mathrm{max}$ steps, then it is halted and the total number of visits to each node $i$ is counted. This number defines $f_i(t=1)$. Then the deposition and the walk is repeated, up to $T$ times, giving the time series $f_i(t)$. We ran simulations with the parameters $M=20000$, $W=100$, $s_\mathrm{max}=100$ and $T=10000$. 

One finds that TFS holds with an exponent $\alpha_\mathrm{T} = 1/2$, see Fig. \ref{fig:Lrwmscaling}. This value is the same as what arises from sums of independent random variables, so the central limit theorem is a possible origin of FS for random walks. The next part presents an analytical calculation that confirms this conjecture.

\addfig{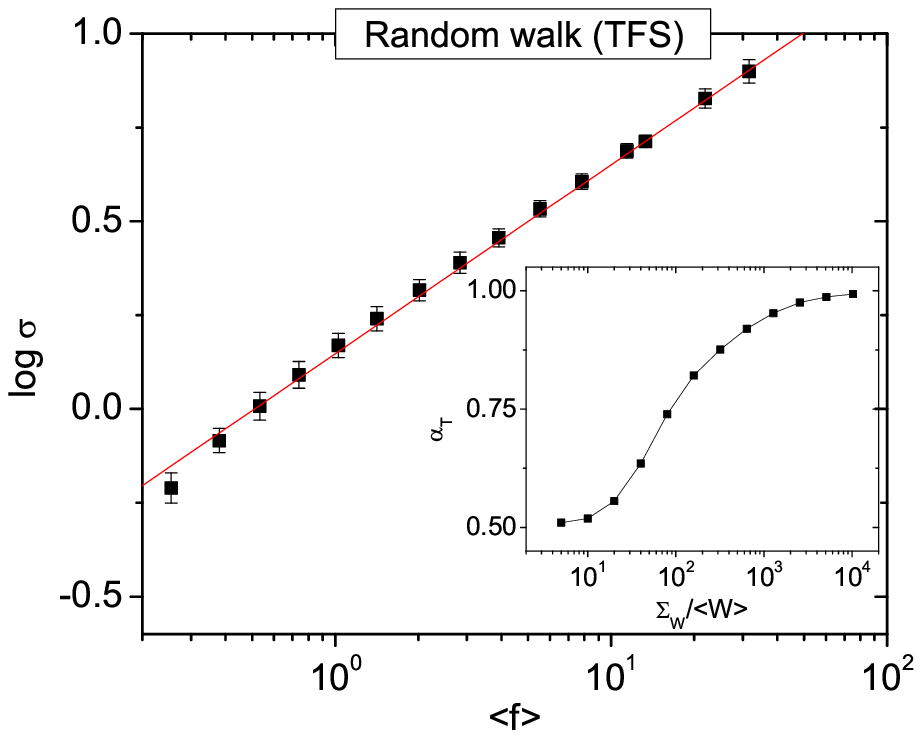}{Fluctuation scaling for the random walker model on the Barab\'asi-Albert network with parameters $M=20000$, $W=100$, $s_\mathrm{max}=100$ and $T=10000$. The fitted exponent is $\alpha_\mathrm{T} = 1/2$ with a little deviation for very small $\ev{f}$. Points were logarithmically binned and $\log \sigma$ was averaged for better visibility, the error bars represent the standard deviations inside the bins. The inset shows the values of the effective exponents as one increases the fluctuations ($\Sigma_W$) in the number of walkers. There is a clear crossover from $\alpha_\mathrm{T} = 1/2$ to $\alpha_\mathrm{T} = 1$.}

\subsubsection{Fluctuation scaling and corrections}
\label{sec:rwmanal}
\label{toc:4.1.2}

The model can be solved based on a master-equation approach \cite{weiss.rw, eisler.internal}. Here we will use elementary probability theory instead. The number of walkers on node $i$ can be calculated from their distribution in the previous time step as:
\begin{equation}
	N_i(s+1) = \sum_{j\in \mathcal K_i}\sum_{n=1}^{N_j(s)} \delta_n(j\rightarrow i; s),
	\label{eq:master}
\end{equation}
where $\delta_n(j\rightarrow i; s)$ is a variable that is $1$ if in step $s$ the $n$'th token was at node $j$ and then it jumped to node $i$ (happens with probability $1/k_j$ to all neighbors of node $i$), and $0$ otherwise. $k_i$ is the degree of node $i$, $\mathcal K_i$ is the set of neighbors of node $i$, and $N_j(s=0)$ corresponds to the initial condition. 

Calculations in Appendix \ref{app:rwm} show that for such a model
\begin{equation}
	\ev{f_i} = s_\mathrm{max}\ev{N_i}=k_i\frac{s_\mathrm{max}W}{\sum_j k_j} = \rho k_i,
	\label{eq:rwmsol1}
\end{equation}
where $\rho = s_\mathrm{max}W/\sum_j k_j$. As the $W$ number of walkers is multiplied by the $s_\mathrm{max}$ and divided by the total number of edges, $\rho$ can be understood as the mean number of walkers passing any edge during the $s_\mathrm{max}$ time steps. Furthermore,
\begin{equation}
	\sigma_i^2 = \sum_{j\in \mathcal K_i} \frac{\sigma_j^2}{k_j^2}+\ev{f_i}.
	\label{eq:rwmsol2}
\end{equation}
The first term on the right hand side is a sum over $k_i$ nodes, but every term is multiplied by $1/k_j^2$, thus they can be neglected to a first order. To a leading order $$\sigma_i^2 = \ev{f_i},$$ thus we find FS with $\alpha_\mathrm{T} = 1/2$. 

\begin{figure*}[ptb]
\centerline{\includegraphics[height=210pt,trim=20 3 50 0]{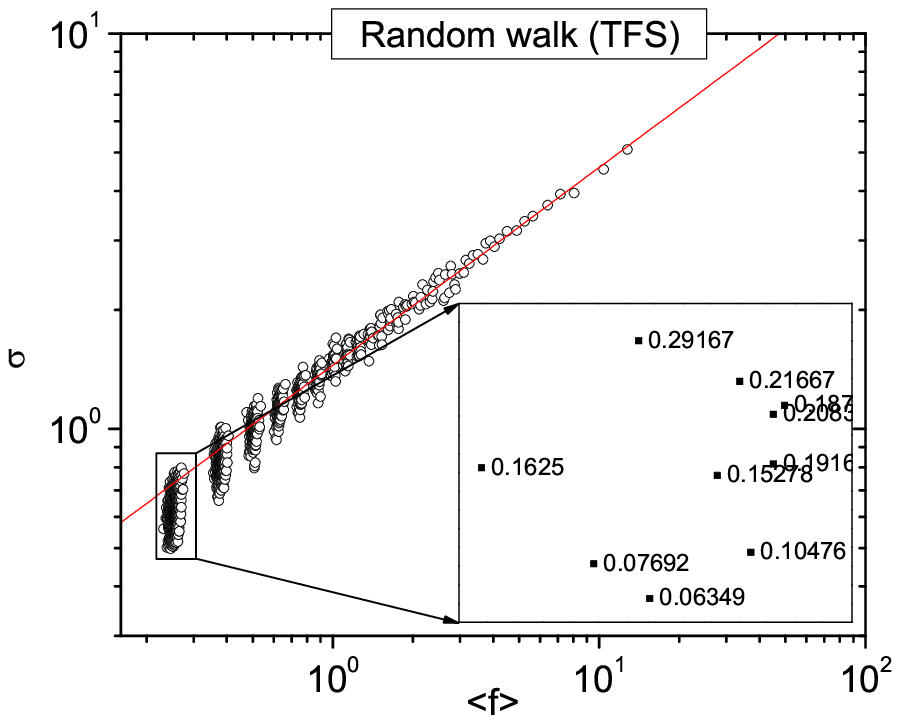}\includegraphics[height=194pt]{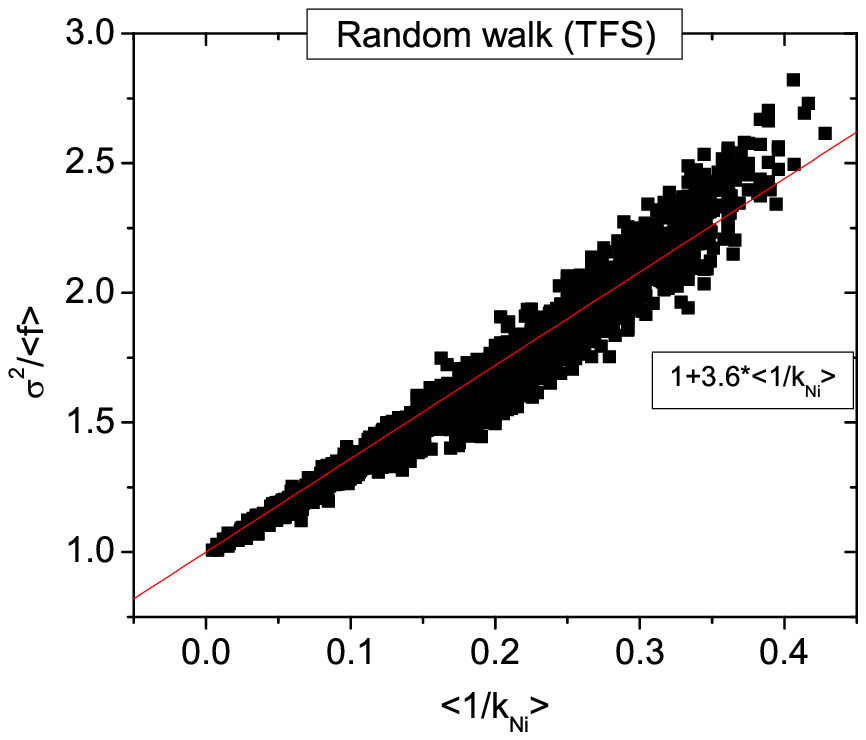}}
\caption{(left) Fluctuation scaling for the random walker model with parameters $M=20000$, $W=100$, $s_\mathrm{max}=100$ and $T=10000$. The same as Fig. \ref{fig:Lrwmscaling}, only without the binning procedure. The scaling law with $\alpha_\mathrm{T} = 1/2$ holds on average, but there is some systematic broadening. The inset shows $10$ randomly selected points from the indicated area, with the average inverse neigh\-bor degree $\ev{1/k_{Ni}}$ indicated for each node. There is a general increasing tendency in $\ev{1/k_{Ni}}$ from bottom to top. (right) The value of $\sigma / \ev{f}^{1/2}$ plotted versus the average inverse neighbor degree $\ev{1/k_{Ni}}$ of the node. There is an approximately linear relationship of the form $\sigma^2/\ev{f} \sim 1 + 3.6\ev{1/k_{Ni}}$.}
\label{fig:Lrwmcorr}
\end{figure*}

The term with the sum presents corrections to the scaling law. Eqs. \eqref{eq:rwmsol1} and \eqref{eq:rwmsol2} could be solved numerically, but to get a qualitative understanding of these corrections it is enough to make a self-consistent solution up to the first non-trivial order. This can be done by taking $\sigma_j^2 = \ev{f_j} = \rho k_j$, and substituting it back into the right hand side of \eqref{eq:rwmsol2}, to find
\begin{equation}
\sigma_i^2 = \sum_{j\in \mathcal K_i}\frac{\rho}{k_j}+\rho k_i = \underbrace{\rho k_i}_{\ev{f_i}} \left (1+a\ev{\frac{1}{k_{Ni}}}\right ),
\label{eq:rwmcorrections}
\end{equation}
where $\ev{1/k_{Ni}}$ is the average inverse neighbor degree of node $i$ and $a=1$. Simulation results supporting this argument are shown in Fig. \ref{fig:Lrwmcorr}. We find that this formula accounts for a large part of the corrections to FS, only the coefficient is different, $a \approx 3.6$.

The qualitative picture from the above three equations is the following. For simplicity let us consider $\rho = 1$, when the average number of tokens at a node equals its degree. Thus on average in every step every node transmits one token on each of its edges to its neighbors. Consequently every node receives typically one token on each edge, so again it will have tokens equal to its degree. These tokens arrive independently, thus the variance is proportional to their number, which implies $\alpha_\mathrm{T} = 1/2$. The corrections in Eq. \eqref{eq:rwmcorrections} imply that nodes with relatively higher degree neighbors (smaller $\ev{1/k_{Ni}}$) exhibit lower fluctuations\footnote{The degree dependence of this correction is related to the assortativity of the network \cite{boccaletti.networks}.}. This is because the number of tokens at a neighboring site with smaller degree is smaller, and thus can have larger relative fluctuations. These fluctuations then affect our site via a stronger variation in incoming tokens.

This argument is important, because it tells us that for random walks \emph{fluctuation scaling is only approximately true}. The local topology of the network can give significant corrections which cause a broadening in the scaling plots and which are not simply due to measurement noise. According to Eq. \eqref{eq:rwmcorrections} the size of the correction term depends on the neighborhood of the node. Because $\ev{f_i} \propto k_i$, very large flux nodes also have many neighbors. In an uncorrelated network the term $\ev{1/k_{Ni}}$ will converge to a constant value with growing $k_i$, its node dependence (and thus the broadening it causes) is diminished.

\subsubsection{The role of node-node interactions and a connection with surface growth}
\label{toc:4.1.3}

Previously we have shown that $\ev{f_i} = \rho k_i$, where $\rho$ is the average number of tokens passing any edge during a time step. The fluctuations of the number of visits to node $i$ come from two sources: (i) the number of such initial tokens at the neighbors, (ii) and how many of the tokens at its neighbors continue their walk to node $i$ in the next step. The number of tokens at a node is coupled with the state of its neighbors in the previous step. This effective interaction between neighboring nodes is the origin of the corrections to FS in Eq. \eqref{eq:rwmcorrections}. To prove this, Menezes and Barab\'asi suggest a mean-field model \cite{barabasi.fluct} which eliminates this interaction as follows.

Instead of a direct contact between nodes, let us completely disconnect the network, and connect every node with its original number of edges to a reservoir. In every step ($s=1\dots S$ as in the original model) the reservoir sends $W$ tokens, their destination is chosen randomly between \emph{the edges}. These tokens return to the reservoir in the next step, but simultaneously $W$ new tokens are sent out, etc. It is trivial that in this case fluctuations of the type (ii) are absent: all nodes are neighbors of the reservoir only, which emits the exact same number of tokens every time. Moreover, the distribution of $f_i(t)$ will be Poissonian with mean and variance $\rho k_i$. Thus exactly
\begin{equation}
	\sigma_i = \ev{f_i}^{1/2} = \sqrt{\rho k_i},
\end{equation}
i.e., $\alpha_\mathrm{T} = 1/2$ without any corrections. Moreover, both the network topology and the "randomness" of the walk was completely eliminated. As suggested by Menezes and Barab\'asi \cite{barabasi.fluct}, the remaining model is equivalent to a surface growth problem. Consider a finite one-dimensional lattice with $\sum_i k_i$ sites. At every time step $W$ tokens are deposited on the surface randomly. The Hurst exponent of the resulting surface is equivalent to the $\alpha_\mathrm{T}$ of the non-interacting model [cf. Eq. \eqref{eq:popalpha}].

This example suggests that \emph{FS in the random walker model is a mean-field property}. The interaction between the nodes is only responsible for higher order corrections that do not change the scaling exponent in general. Most models in the literature are either non-interacting in this sense \cite{ballantyne.model, measles, cells} or this interaction is not relevant \cite{barabasi.fluct, eisler.internal}. At most, complex dynamics is limited to the structure \emph{within} \cite{kilpatrick.ives} the nodes, but not between the nodes. There exist a few studies of transport models on complex networks where the interaction between the nodes becomes relevant. In these models fluctuation scaling breaks down and topology-dependent crossovers appear due to congestion \cite{tadic.loops} or the presence of multiplicative noise \cite{menezes.flux}.

\subsubsection{The role of external driving}
\label{toc:4.1.4}
Finally, let us briefly remark on the behavior of the model in the presence of external driving. One can allow the number $W$ of walkers to fluctuate between the times $t$ as $$W(t) = \ev{W}+\Sigma_W \times G(t).$$ We chose $G(t)$ as i.i.d. standard Gaussians, but the findings are largely independent of the shape of the distribution. If at any time $W(t)$ became less than zero, we set it $W(t)=0$. If we restrict ourselves to the mean-field solution, then at every node \cite{barabasi.fluct} 
\begin{equation}
\sigma^2_i = \ev{f_i}+\left[\frac{\Sigma_W}{\ev{W}}\right]^2\ev{f_i}^2.
\label{eq:sigmaW}
\end{equation}
This result implies that when $\Sigma_W > 0$, there is a crossover from $\alpha_\mathrm{T} = 1/2$ to $\alpha_\mathrm{T} = 1$ around the node strength $\ev{f} \sim \ev{W}^2/\Sigma^2_W$.

The process was simulated with the other parameters set as before. With the increase of $\Sigma_W$ the best fit to Eq. \eqref{eq:taylori} yields intermediate effective exponents between $1/2$ and $1$, see the inset of Fig. \ref{fig:Lrwmscaling}. However, these are not "true" exponents, only signatures of the crossover.

\subsubsection{Impact inhomogeneity}
\label{toc:4.1.5}

The random walker model can be modified \cite{eisler.internal} to entail Eq. \eqref{eq:beta}. This means that when a walker steps onto a site with typically more visitations, it generates a higher impact. Since the number of visits is proportional to the degree of the node ($\ev{N_i}\propto k_i$), in order to have the impact inhomogeneity relationship $\ev{V_i}\propto\ev{N_i}^\beta$, one can simply introduce that for a token visiting a node of degree $k_i$, the impact should be $\ev{V_i}=k_i^\beta$. Simulation results perfectly conform with the theory, $\alpha_\mathrm{T}(\Sigma_W = 0)$ depends on $\beta$ as expected from Eq. \eqref{eq:alphabeta}. The crossover persists to $\alpha_\mathrm{T} = 1$ when one introduces a large variation in the number of tokens, see Fig. \ref{fig:Grwmbeta}.

\addfig{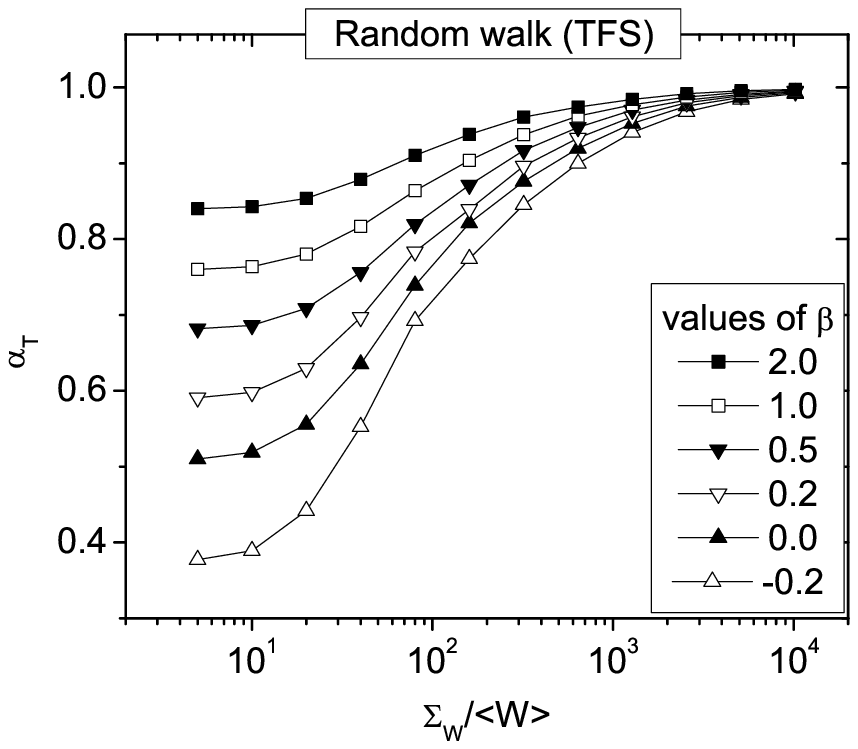}{Fluctuation scaling for the random walker model with inhomogeneous impact, parameters $M=20000$, $W=100$, $s_\mathrm{max}=100$, $T=10000$ and various values of $\beta$. The case $\beta = 0$ is the same curve as in the inset of Fig. \ref{fig:Lrwmscaling}. The $\alpha_\mathrm{T}(\Sigma_W\rightarrow 0)$ limit is well described by Eq. \eqref{eq:alphabeta}. For $\Sigma_W \gg \ev{W}$ every system displays a crossover to $\alpha_\mathrm{T} = 1$.}

\subsection{Critical fluctuations and finite size scaling}
\label{sec:ising}
\label{toc:4.2}

The mechanism how (spatial) correlations produce non-trivial values of $\alpha$ draws on some fundamental knowledge in statistical physics. Critical systems are known to exhibit anomalous fluctuations due to the presence of strong, but non-trivial correlations. These originate from the interactions of the internal constituents as for e.g. Ising spins.

It is instructive to consider the simple ferromagnetic case, like the nearest-neighbor Ising model \cite{reichl} on a $d$ dimensional square lattice. {Because this model does not \emph{a priori} have dynamics, its analysis can be understood in the language of ensemble averages. 

The number of spins is $N=L^d$, where $L$ is the linear size of the lattice. At the critical point local magnetization has a diverging correlation length, and the correlation function becomes of the power law form
\begin{equation}
	C(\mathbf r) \propto \frac{1}{r^{d-2+\eta}}.
	\label{eq:isingcorr}
\end{equation}
The squared fluctuations of total magnetization [$\overline \sigma^2(M_L)$] are known to be proportional to the susceptibility $\chi$, 
and for finite systems both quantities diverge at the critical point as
\begin{equation}
	\evl{\sigma^2(M_L)} \propto \chi \propto L^{d+\gamma/\nu}.
	\label{eq:chi1}
\end{equation}
This is one of the well known results of finite size scaling (FSS) \cite{cardy}.

The susceptibility can be calculated as the integral of the correlation function:
\begin{equation}
	\chi = \frac{N}{k_BT} \int d^d\mathbf rC(\mathbf r) \propto N \int^L \frac{d^d\mathbf r}{r^{d-2+\eta}} \propto L^{d+2-\eta}.
	\label{eq:chi2}
\end{equation}
{It is well known that the exponents in Eqs. \eqref{eq:chi1} and \eqref{eq:chi2} are related, $\gamma/\nu = 2-\eta$ (Fisher's law \cite{reichl}). At the critical point, due to the interactions between the spins, the susceptibility becomes super-extensive, i.e. it grows faster than $\propto L^d$, a typical sign of criticality.}

Let us now consider an ensemble of finite Ising systems at the critical temperature with zero external field and with various linear sizes, and let the signal be the $N_{L,\uparrow}$ number of "up" spins. Of course the total number of up and down spins is constant: $$N_{L,\uparrow}+N_{L,\downarrow}=L^d,$$ and their difference gives the magnetization as $$M_L=N_{L,\uparrow}-N_{L,\downarrow}.$$ With the notation $o(L^p)/L^p \rightarrow 0$, at the critical point $$\overline{N_{L,\uparrow}} = L^d/2 + o(L^d).$$ On the other hand, the fluctuations of $M$ and $N_{\uparrow}$ are proportional, because $$M_L=2N_{L,\uparrow}-L^d,$$ and so $$\overline{\sigma(N_{L,\uparrow})}^2=\overline{\sigma(M_L)}^2/2 \propto L^{d+2-\eta} + o(L^{d+2-\eta}).$$ Consequently, to a leading order, there exists EFS between the fluctuations and the mean of the number of up spins:
$$\overline{\sigma(N_{L,\uparrow})}^2 \propto \overline{N_{L,\uparrow}}^{2\alpha_\mathrm{E}}$$
with
\begin{equation}
	\alpha_\mathrm{E} = \frac{1}{2}+\frac{2-\eta}{2d} = \frac{1}{2}+\frac{\gamma/\nu}{2d}.
	\label{eq:isingalpha}
\end{equation}

The above are true up to the upper critical dimension, which is $d_c = 4$ for the Ising model \cite{cardy}. The mean-field results can be recovered by substituting the corresponding values: $d=d_c = 4$, $\gamma_\mathrm{MF}=1$, $\nu_\mathrm{MF}=1/2$, $\eta_\mathrm{MF}=0$. Finally $\alpha_{\mathrm{E,MF}} = 3/4$, in agreement with the direct mathematical proof of Ellis and Newman \cite{ellis.cw}. {Moreover, at the critical point the susceptibility is superextensive, so $\chi$ must grow faster than $L^d$. This means that in Eq. \eqref{eq:chi2} $d+2-\eta > d$, and thus $\eta < 2$. On the other hand if $\eta$ is non-negative, then from the constraints $0\leq\eta < 2$ and $d\geq 1$ it immediately follows that $1/2\leq\alpha_\mathrm{E} < 1$. This range is also valid for the analogous behavior of all $n$-vector models.}

This result is two-fold, depending on how we look at it:
\begin{enumerate}[(i)]
	\item The exponent $\alpha_\mathrm{E}$ resembles the finite-size scaling exponent of fluctuations/susceptibility. Thus in this case \emph{fluctuation scaling essentially finite-size scaling.} The difference is that the FS calculation can be done even when there is no data available about "system size". Instead, because $N_{\uparrow}$ is a positive extensive quantity, we know that its expectation value will be proportional to the system size, and thus it can act as a surrogate variable for $L^d$. An anomalous value of the FS exponent can be related to critical behavior, although -- as previous sections suggest -- not necessarily. We will discuss this question in detail in Section \ref{sec:scaling}.
	\item The finding that when the constituents are long-range correlated gives rise to anomalous values of $\alpha$, leaves us with a recipe how to construct simple models that display $1/2<\alpha<1$. The simplest scenario is described in detail in Sections \ref{sec:sync} and \ref{sec:binary}.
\end{enumerate}

What is the case with $N_\uparrow$ off the critical point? In the paramagnetic phase the mean number of up spins is exactly $N/2$, while the fluctuations are of order $N^{1/2}$, thus $\alpha_\mathrm{E} = 1/2$. The ferromagnetic case is a more delicate issue, because the infinite system is not ergodic: spontaneous magnetization is 
symmetry breaking. For finite systems with a local (e.g. Glauber) dynamics it takes a finite (but very long) time for magnetization to change direction. The phenomenon is more easily interpreted via an (unrestricted) ensemble of equilibrium ferromagnets. Here still $\overline f = N/2$, because configurations magnetized up and down average out. The fluctuations on the other hand are macroscopic, $\evl{\sigma_L} \propto 2\evl{\vert M\vert} \propto L^d$. Thus $\alpha_\mathrm{E} = 1$. In sum, the paramagnet-ferromagnet phase transition is signaled by FS as an abrupt change between the two universality classes (similarly to the Satake-Iwasa forest model \cite{satake.iwasa, ballantyne.model}). At the critical point one finds intermediate exponents that can be calculated from the usual critical exponents. However, it is of fundamental importance that the anomalous FS is not observed in the order parameter $M$. Instead, it is observed in an extensive quantity, and only whose \emph{fluctuations} reflect the anomalous fluctuations of the order parameter. FS is there in $M$, but with a trivial exponent: From finite size scaling $\overline M \propto L^{d-\beta/\nu}$. This, combined with Eq. \eqref{eq:chi2} leads to $\overline{\sigma(M_L)} \propto M^{(d+\gamma/\nu)/[2(d-\beta/\nu)]}$. Due to the hyperscaling relation $\gamma + 2\beta = d\nu$ this means $\alpha_E = 1$.


{The critical point is a very special state of a system, while fluctuation scaling with $1/2<\alpha<1$ occurs very often. To make criticality a viable explanation for these non-trivial values of $\alpha$ it is important to notice that certain types of dynamics under strong external driving can self-organize to their critical state without the fine-tuning of any parameters \cite{bak.soc, bak.book, paczuski.solar}. Many real life systems display the classical signs of \emph{self-organized criticality} (such as power-law distributions, long-range correlations, etc.) and the value of $\alpha$ can help to understand the dynamical origins of these observations.}

\subsection{Scaling and multiscaling}
\label{sec:scaling}
\label{toc:4.3}

Scaling has a fundamental importance in statistical physics. It has found countless successful applications starting with critical phenomena \cite{stanley.phase}, but more recently also outside the classical domain of physics, for example in ecology \cite{banavar.ecology}. In many cases scaling is not bound to a specific set of system parameters like in the case of critical phenomena, but it is the generic behavior of the system as in polymers \cite{degennes.book}, surface growth \cite{barabasi.stanley.book} and self-organized criticality \cite{bak.soc}. Mono-scaling or gap scaling means that the probability distribution of a quantity $f$ depends on the parameter $L$, usually the system size, as
\begin{equation}
	\mathbb P(f, L) = f^{-1}F\left (\frac{f}{L^{\Phi}}\right),
	\label{eq:gapscaling}
\end{equation}
where $F$ is a scaling function and $\Phi$ is some constant. This form can account for a number of observations about power law behavior in real systems.

Both gap scaling and fluctuation scaling characterize a large number of complex systems. Nevertheless, for the same quantity \emph{only one} can be true except in a special case: If a quantity shows both gap scaling and fluctuation scaling, then this automatically implies $\alpha = 1$. One can reverse this argument: If for a quantity one finds fluctuation scaling with $\alpha < 1$ then it cannot exhibit gap scaling.

The proof is straightforward. Any moment of $f$ can be calculated as
\begin{eqnarray}
\evl{f^q_L} = \int_{f_0}^\infty df f^q \mathbb P(f, L) \simeq K_qL^{q\Phi},
\label{eq:nq}
\end{eqnarray}
where "$\simeq$" denotes asymptotic equality and $K_q > 0$. From Eq. \eqref{eq:nq} it follows that
\begin{eqnarray}
	\evl{\sigma^2_L} = \evl{f^2_L}-\evl{f_L}^2 \simeq \nonumber \\ K_2 L^{2\Phi}-K_1^2 L^{2\Phi}=(K_2-K_1^2)L^{2\Phi}.
\label{eq:sigma}
\end{eqnarray}
We combine EFS and Eq. \eqref{eq:sigma}, eliminate $L$ and find that now $\evl{\sigma^2_L} \propto\evl{f_L}^2$, i.e., $\alpha = 1$. 

The only possibility for the coexistence of gap scaling \eqref{eq:gapscaling} and fluctuation scaling \eqref{eq:taylorN} with $\alpha < 1$ is when the constant factor in the variance vanishes: $$(K_2-K_1^2)=0.$$ In this case the gap scaling form does not describe the variance, that is instead given by the next order (correction) terms. Nevertheless, even if it is so, the \emph{leading} order of the variance is still zero, and consequently $F$ is proportional to a Dirac-delta: $$F\left(\frac{f}{L^{\Phi}}\right) \propto \delta\left(f/L^{\Phi}-K_1\right).$$ This case is pathological, and it is usually not considered as scaling. In fact, the previous section contained one such example: The number of up spins in a critical Ising model follows this sort of statistics. Fluctuations scale anomalously ($\evl{\sigma^2_L} \propto L^{d+2-\eta}$), whereas their leading order vanishes because $\evl{f^2} \simeq \evl{f}^2 \simeq L^{2d}/4$. Such strange scaling arises as a sign of criticality when the scaling variable is an extensive quantity, for which only the fluctuations are connected to those of the order parameter.

For example, in ecology there do exist species with $\alpha \approx 1$, for which a gap scaling form of the probability density of $f$ could be valid. However, this value is by no means universal (cf. Fig. \ref{fig:Gtaylorshifted}). Similarly, $\alpha < 1$ was observed for Internet router traffic \cite{barabasi.fluct} or the traded value on stock markets \cite{eisler.non-universality}. These quantities cannot have a gap scaling form.

{Instead of gap scaling, one can assume multiscaling, but the results do not change crucially. A probability distribution shows \emph{multiscaling} if its size dependence is of the form
\begin{equation}
	\ln \mathbb P(f, L)/\ln (L/L_0) = -F[\ln(f/f_0)/\ln(L/L_0)],
	\label{eq:multiscaling}
\end{equation}
where $f_0$ and $L_0$ are appropriately chosen constants. The moments can be calculated by
expressing the density function from Eq. \eqref{eq:multiscaling} and substituting into the definition
\begin{eqnarray}
\overline{f_L^q}=\int_0^\infty f^q \mathbb P(f, L)df = \nonumber \\ 
\int_0^\infty f^q \left(\frac{L}{L_0}\right)^{-F[\ln(f/f_0)/\ln(L/L_0)]}df \simeq \nonumber \\
f_0^q\left(\frac{L}{L_0}\right)^{qa(q)}\left(\frac{L}{L_0}\right)^{-F[a(q)]} \simeq K_q L^{\tau(q)}.\nonumber
\end{eqnarray}
The usual approach is that the value of the integral is dominated by the point $f_*(q)$ where the integrand is maximal. Then $$a(q)=\frac{\ln [f_*(q)/f_0]}{\ln (L/L_0)},$$
and $$\tau(q) = \max_a [qa-F(a)],$$ or equivalently $\frac{\tau(q)}{\partial q} = a.$ Now we are back at the same situation as with gap scaling, since $$\overline{f_L} \simeq K_1 L^{\tau(1)}$$ and $$\evl{\sigma_L^2} = \overline{f_L^2}-\overline{f_L}^2 \simeq K_2L^{\tau(2)}-K_1^2L^{2\tau(1)}.$$ One expects that $\tau(2)\geq 2\tau(1)$, because the variance must remain non-negative for arbitrarily large $L$. If $\tau(2) > 2\tau(1)$ then the first term dominates $\evl{\sigma_L^2}$, and $\alpha = \frac{\tau(2)}{2\tau(1)},$ but this value is greater than $1$. For example Tebaldi et al. \cite{stella.btw} report that in the Bak-Tang-Wiesenfeld sandpile model of $L$ linear size, the distribution of the number of topplings $f$ in an avalanche follows $\evl{f^q_L} \propto L^{\tau(q)}$ with $\tau(1) \approx 2$, and $\tau(2) \approx 4.7$. This results in an $\alpha \approx 1.17$.


The other possibility is again $\tau(2)=2\tau(1)$, and $\alpha = 1$ (unless the leading order terms in $\sigma^2$ compensate to zero). This solution offers nothing new compared to gap scaling. Such relationships can be seen, e.g., in the very same BTW model for the distribution of the area affected by avalanches \cite{stella.btw}.
%
%

The conclusion: If a quantity shows gap scaling with a scaling function which is not fully degenerate (not a Dirac-delta), it must follow $\alpha = 1$. If there is multiscaling, then fluctuation scaling with $\alpha > 1$ is also possible, but such values are rarely observed and should be taken with care.


%


%
%

\subsection{Binary forest model}
\label{sec:binary}
\label{toc:4.4}

In this section we introduce a toy model that can be used to better illustrate the ideas of Sections \ref{sec:ising}-\ref{sec:scaling}. Moreover, we will show that those are in full analogy with the findings of Section \ref{sec:timeecol} for the reproductive activity of trees. For an easier understanding we will present the model in that language.

Let us consider a forest that consists of $N$ trees. For simplicity we also assume that these are situated on a one-dimensional regular lattice, but any higher dimensional generalization is straightforward. In the year $t$ the reproductive activity (i.e. seed count) of every tree $n$ is characterized by a random variable $V_n(t)$. Again, for simplicity we consider $V$'s as binary variables, which are $1$ with probability $p$ and $0$ with probability $1-p$. Because it takes several years for a new tree to reach its full reproductive capabilities, given that the observation period is short enough, we can neglect the changes in $N$ due to seed production and tree growth.

The year-to-year correlations in seed counts are neglected. On the other hand, it is known \cite{koenig.masting} that the reproductive activity of forests exhibits long-range \emph{spatial} dependence, with significant positive correlations for distances of thousands of kilometers. The distance dependence can be fitted approximately by
\begin{equation*}
	C(\Delta n) \propto \ev{V_{n}V_{n+\Delta n}}-\ev{V_{n}}^2 \propto \Delta n^{2H_V-2} \tag{\ref{eq:Cdn}},
\end{equation*}
see Section \ref{sec:timeecol}. The total seed count is given by the usual form
$$f_N = \sum_{n=1}^NV_n.$$
The standard deviation of the sum of random variables correlated according to Eq. \eqref{eq:Cdn} scales as 
$$\sigma_N = \sqrt{\ev{f_N^2}-\ev{f_N}^2}\propto N^{H_V}$$ with $H_V$ being the Hurst exponent [cf. Eq. \eqref{eq:tickhurst}], whereas $$\ev{f_N} = pN.$$ The two equations can be combined into TFS with
\begin{equation*}
	\alpha_\mathrm{T} = H_V. \tag{\ref{eq:popalpha}}
\end{equation*}

To the careful reader it should be clear that almost the same model was discussed in Section \ref{sec:ising}. There we argued that in a critical Ising model whether any given spin points upwards ($1$) or downwards ($0$) is essentially a binary random variable with $p=1/2$. Moreover the spin alignments are power-law correlated in space, such that the power of the decay is related to the FS exponent $\alpha$. This was expressed by Eq. \eqref{eq:isingalpha}, which is essentially equivalent to \eqref{eq:popalpha}. The binary forest model only differs from the Ising case in that correlations between the random variables are given \emph{a priori}, and not generated by the thermodynamics.

Now we can move on to simulation results.\footnote{Except for the trivial case $H_V = 1/2$ (when $V$'s are not strongly correlated) the above model is not very straightforward to simulate. We generated a one-dimensional fractional Brownian motion time series by applying the method of Koutsoyiannis \cite{koutsoyiannis.hurst}, and then converted it into a sequence of $0$'s and $1$'s\footnote{For higher dimensions it is necessary to use a more refined method, for example the one introduced by Prakash et al. \cite{havlin.percolation} for the simulation of site percolation on long-range correlated lattices.}. The conversion slightly decreases the value of the Hurst exponent, which thus had to be measured independently by Detrended Fluctuation Analysis \cite{dfa}. For simplicity, we fixed the number of trees, because the effect of externally imposed noise ($\Sigma^2_N>0$) has already been studied in detail in Section \ref{sec:alpha1} and, e.g., Refs. \cite{barabasi.fluct, eisler.internal} for other models.} Fig. \ref{fig:Gpopdfa} shows the dependence of $\sigma_N$ on $N$ and $\ev{f}$, the two plots are basically equivalent due to $\ev{f}=pN$. Fig. \ref{fig:GpopexampleNt}(left) illustrates that the fluctuations in systems of the same size increase rapidly with $H_V$. This is due to a strong synchronization of the individual constituents [see Fig. \ref{fig:GpopexampleNt}(right)]. The relationship \eqref{eq:popalpha} is illustrated in Fig. \ref{fig:Gpopalphadfa}.

\addtwofigs{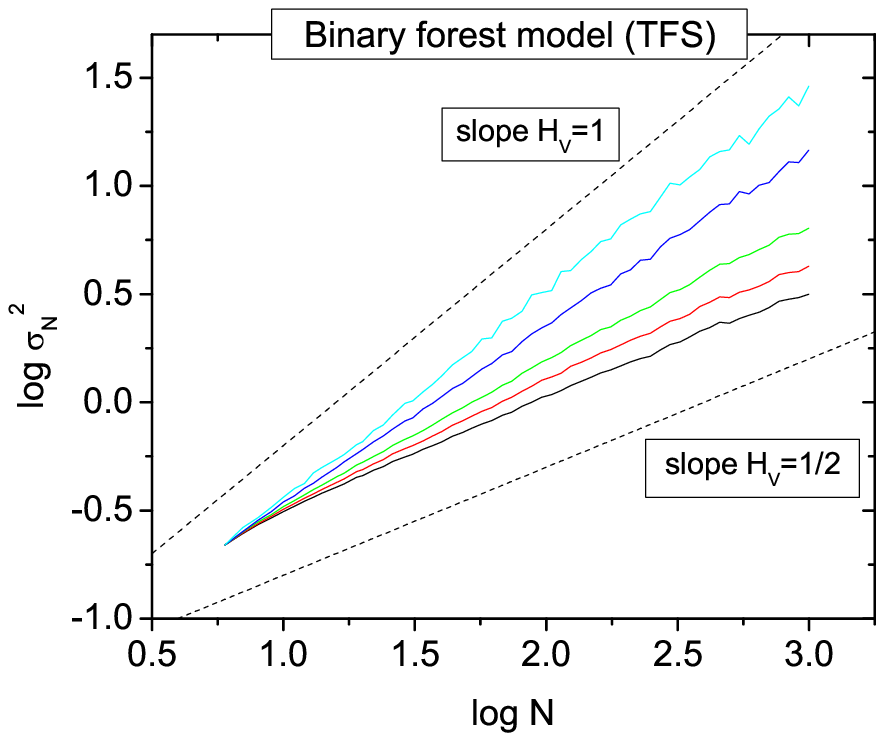}{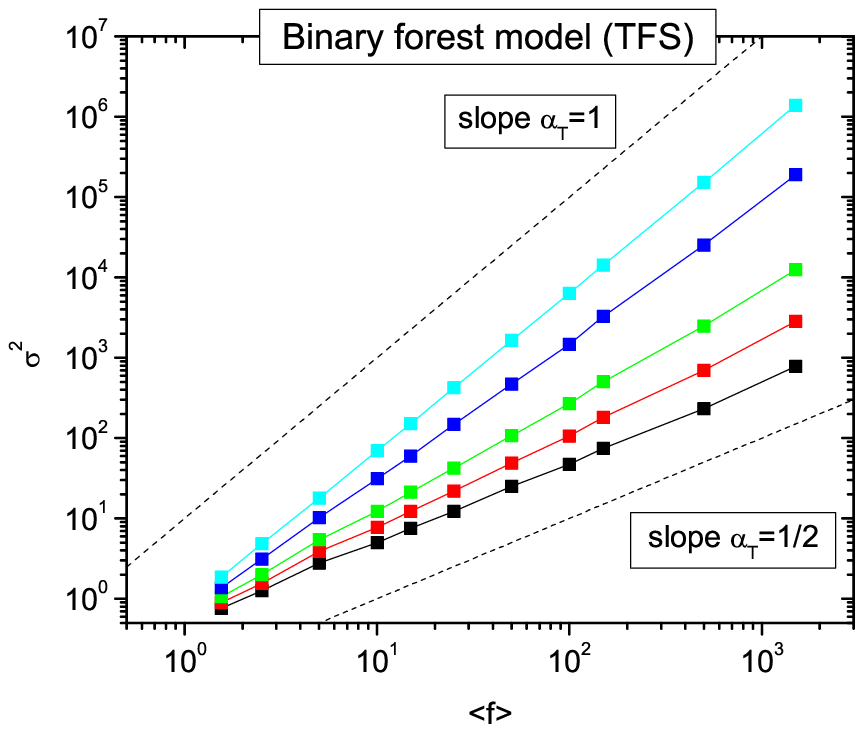}{(left) Scaling plots of $\sigma$ versus $N$ generated by Detrended Fluctuation Analysis of $V_n(t)$ in the binary forest model. The slopes correspond to the (spatial) Hurst exponents $H_V\approx 0.5,\dots,0.95$ from bottom to top, see Eq. \eqref{eq:tickhurst}. (right) Scaling plots $\sigma$ versus $\ev{f}$ for FS in the same data. The slopes correspond to the values of $\alpha_\mathrm{T} \approx 0.5,\dots,0.95$.}
\addtwofigs{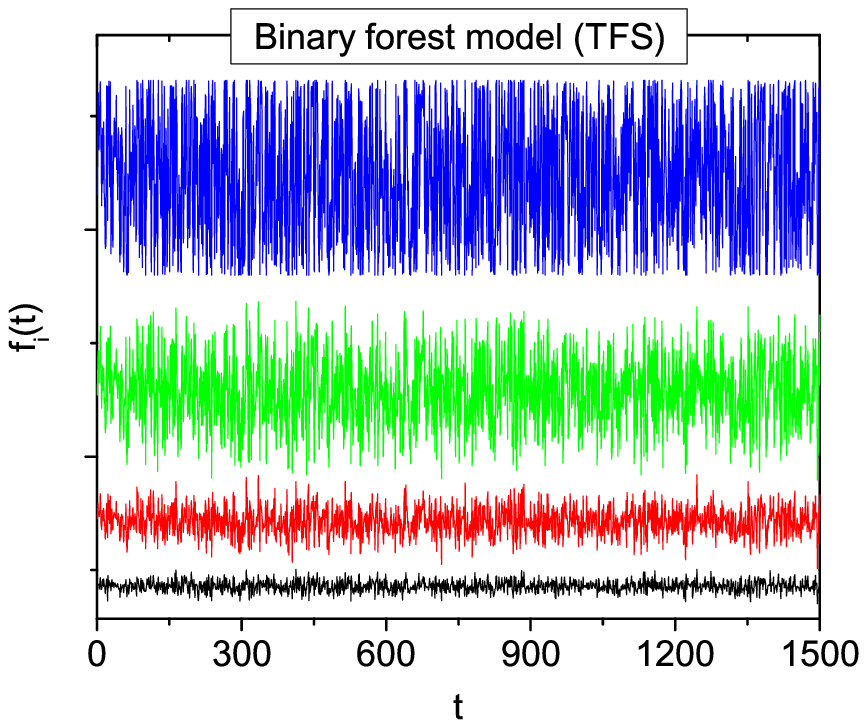}{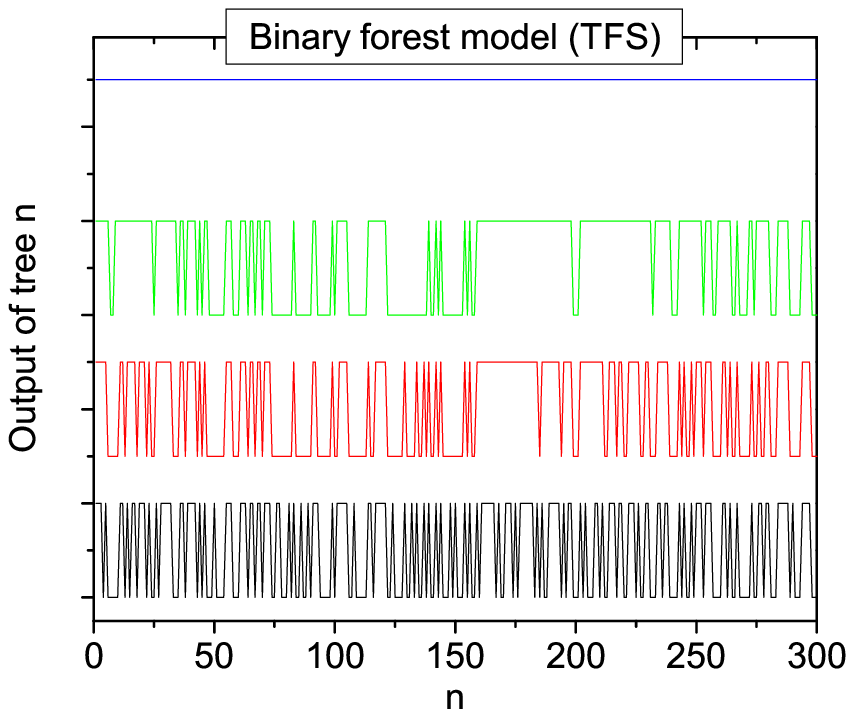}{(left) Examples of $f_i(t)$ time series for a "forest" with $N=300$ "trees". The Hurst exponent $H_V$ between the trees was varied: $H_V \approx 0.5, 0.65, 0.8, 0.95$ increasing from bottom to top. The data were shifted by the addition of a constant, but they were not stretched in any way. One can see that due to the increasing synchronization of the constituents, relative fluctuations increase rapidly. (right) Snapshot of $V_{i,n}$ series (at a fixed time $t$) for a forest with $N=300$ elements. The data were shifted by the addition of a constant. The Hurst exponent $H_V$ between the elements was varied: $H_V \approx 0.5, 0.65, 0.8, 0.95$ increasing from bottom to top. Spatial synchronization increases with the growth of the Hurst exponent.}
\addfig{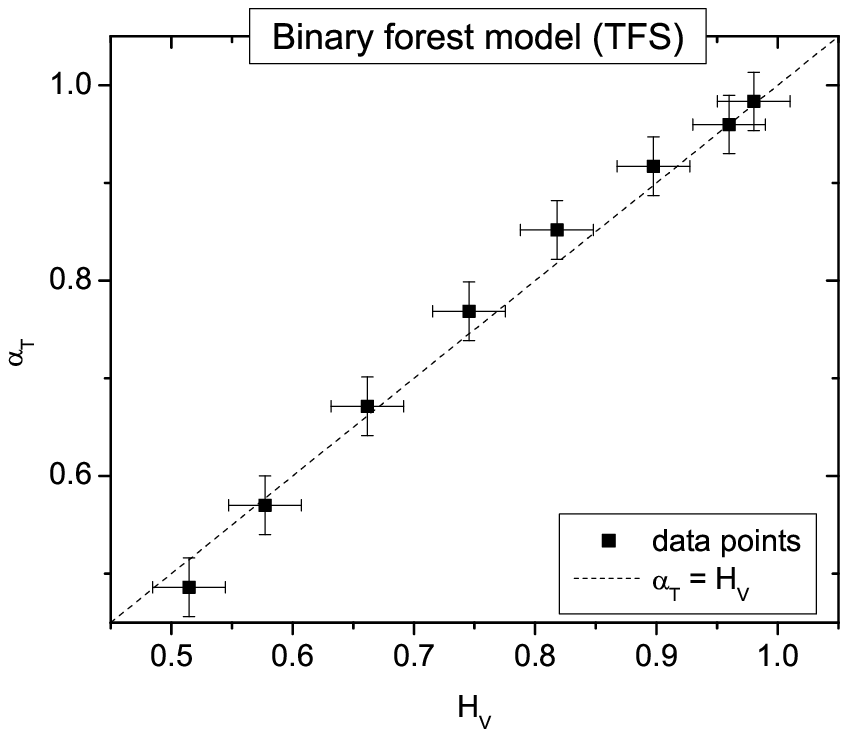}{The equality $\alpha_\mathrm{E} = H_V$ (dotted line) in simulations of the binary forest model. The measurement points align very closely to the line, with some statistical deviations.}

\section{Discussion}
\label{sec:disc}
\label{toc:5}

In this section we present our view about unsettled questions related to fluctuation scaling. We also discuss recent, sometimes controversial techniques that might help the understanding of FS.

\subsection{Separation of global and local dynamics}
\label{toc:5.1}

In Section \ref{sec:univcl} we argued that a system whose internal dynamics can be mapped onto the central limit theorem displays fluctuation scaling with $\alpha = 1/2$. On the other hand, if one imposes a strong external driving to the system, the behavior crosses over to $\alpha = 1$. One example was shown in Section \ref{sec:rw} in the case of random walks on complex networks. There the fluctuation was given by Eq. \eqref{eq:sigmaW}, which has the structure
\begin{equation}
\sigma^2_i = \ev{f_i}+A^2\ev{f_i}^2,
\label{eq:A}
\end{equation}
where $A$ is proportional to the strength of the external driving. If $A \ll 1$ one finds $\alpha = 1/2$, whereas in the strongly driven limit $A \gg 1$ the first term is negligible and $\alpha = 1$.

Now assume that we do not know the strength of external driving and we want to approximate it from data. We can introduce the global activity $F(t)$ of the system as a sum over all constituents:
\begin{equation}
	F(t) = \sum_{i=1}^N f_i(t).
	\label{eq:F}
\end{equation}
Ref. \cite{barabasi.separating} suggests that if our system has many elements, then $F(t)$ will be proportional to the external force, because the independent fluctuations of the elements average out in the sum \eqref{eq:F}, and what remains is only the factor of the common external driving. This argument implicitly assumes, that the external force contributes to the fluctuations of the elements in a \emph{coherent} way, i.e., $f_i(t)$ can be written in the form
\begin{equation}
	f_i(t) = f_i^\mathrm{int}(t) + f_i^\mathrm{ext}(t),
	\label{eq:Ai}
\end{equation}
where 
\begin{equation}
	f_i^\mathrm{ext}(t) = A_i F(t).
	\label{eq:fiext}
\end{equation}
This formula is a form of linear response, where: (i) $A_i$ is not allowed to depend on time because of stationarity. (ii) More importantly, all nodes are affected by driving instantaneously or with the same constant time lag.

After the summation of Eq. \eqref{eq:F} we find that it is consistent with Eq. \eqref{eq:Ai}, if the normalization condition $\sum_i A_i = 1$ is satisfied. In order to keep Eqs. \eqref{eq:A} and \eqref{eq:Ai} consistent in the strongly driven limit, the only possible choice is
\begin{equation}
	A_i = \frac{\ev{f_i}}{\ev{F}}.
	\label{eq:Ai2}
\end{equation}
By this definition automatically $\ev{f_i^\mathrm{int}} = 0$ and $\ev{f_i^\mathrm{ext}} = \ev{f_i}$. All time series have finite standard deviations, which are defined in the usual way, for example $\sigma_F = \sqrt{\ev{F^2}-\ev{F^2}}$. With these  
\begin{equation}
\sigma_i^\mathrm{ext} = \frac{\sigma_F}{\ev{F}} \ev{f_i}, 
\end{equation}
so the external component follows FS with $\alpha = 1$ in \emph{any} system. This appears consistent with the fact that in strongly driven systems $f$ itself also shows $\alpha = 1$, not only the external component. However, this is in fact just a trivial consequence of how the external component was defined.

Ref. \cite{barabasi.separating} calls the process of assigning the internal and external components "noise separation", and claims that the procedure works well for the random walk model. For each node they define a noise ratio $$\eta_i = \frac{\sigma_i^\mathrm{ext}}{\sigma_i^\mathrm{int}},$$
which is zero in the absence of external driving, and large when the fluctuations of the external component are dominant. 

It would be tempting to attribute the real world observations of $\alpha \approx 1$ to external driving, and show that in these cases typically $\eta \gg 1$. However, we will demonstrate on some examples that noise separation has strong limitations.

Ref. \cite{barabasi.separating} finds that the fluctuations of Internet (Abiline backbone) traffic show TFS with $\alpha = 1/2$ and typically $\eta \sim 0.1$. While this appears very convincing, a more detailed analysis of a subset of the same data \cite{duch.internet} instead finds $\alpha = 0.7-0.8$. The latter study suggests congestion as the origin of the increase value of $\alpha$, and does not assume any external driving force.

Ref. \cite{barabasi.fluct} reports, that river level fluctuations fall into the class $\alpha = 1$. It seems plausible that water levels fluctuate due to rainfall on the river basin, which can be understood as external driving. However, noise separation is impossible here, because the driving is not coherent. The global factor $F(t) = \sum_i f_i(t)$ is meaningless, because the response times of the water level, and the timing of precipitation vary from river to river. Hence Eq. \eqref{eq:fiext} is not valid.

To our knowledge, our study \cite{eisler.non-universality} was the first to reveal fluctuation scaling in the trading activity of stocks. Noise separation was carried out there, finding that the typical value of $\eta_i$ increases with the time window $\Delta t$. Because $\alpha$ also shows a similar tendency (cf. Fig. \ref{fig:Lstock}), we suggested that external driving must play a role in this effect. We also argued that this is because information needs a finite time to spread on the market. On the scale of a few minutes the role of external information is small and localized, whereas on the long run trading is dominated by the external macroeconomic trends and news.

Later we proposed a much simpler explanation \cite{eisler.unified}, which was also summarized in Sections \ref{sec:stock} and \ref{sec:gamma} of this review. In the stock market (and human dynamics, see Section \ref{sec:human}) one observes, that for long times $\alpha(\Delta t) = \alpha^* + \gamma \log \Delta t$ and $H_i = H^* + \gamma\log\ev{f_i}$ with some $\gamma > 0$. These laws have not yet been related to any external force, even though the possibility cannot be ruled out.

How would noise separation work in this case? 
\begin{enumerate}
\item Clearly $\sigma_F \propto \Delta t^{H_F}$, with $H_F \approx \max_i H_i$, because $F$ is the sum of all $f_i$'s, and the scaling of the sum is dominated by the highest Hurst exponent.  
\item $\sigma_i^\mathrm{ext} \propto \sigma_F \propto \Delta t^{H_F}$. On the other hand, if $\sigma_i^\mathrm{ext} < \sigma_i^\mathrm{int}$, then one expects that qualitatively $\sigma_i^\mathrm{int} \propto \Delta t^{H_i}$.
\item Thus the ratio $\eta_i = \sigma_i^\mathrm{ext}/\sigma_i^\mathrm{int}$ should typically grow as long as $\eta_i < 1$. This observation of Ref. \cite{eisler.non-universality} is hence no proof of any particular external influence.
\end{enumerate}

While to present further calculations is not the purpose of this review, we believe, that $\eta \simeq 1$ can arise from spurious effects. A value $\eta_i > 1$ consistently, for many nodes has only been observed in a single study where $\eta \simeq 1.5$ \cite{jiang.fluxes}. Our present understanding is that noise separation has a limited range of applicability. 

Finally, we would like to point out that to identify the ensemble average \eqref{eq:F} with some external force is somewhat controversial. Since we do not have any information about the origin or the physical meaning of the factor $F(t)$, it is probably more appropriate to call this and $f_i^\mathrm{ext}$'s \emph{global} and not \emph{external} factors. Accordingly, $f_i^\mathrm{int}$'s are better called \emph{local}, rather than \emph{internal} factors when it is unknown how much they represent internal processes at the nodes. 

\subsection{Limit theorems for sums of random variables}
\label{sec:limit}
\label{toc:5.2}

In Section \ref{sec:alpha1/2} we briefly remarked on the connection of $\alpha = 1/2$ to the central limit theorem. We recall that $f$ is written as a sum over the constituents (other random variables) whose number $N$ we will consider as time independent: $$f = \sum_{n=1}^{N}V_n.$$ Let us assume that a general form of central limit theorem is applicable, so that 
\begin{equation}
	\frac{\sum_{n=1}^NV_n-N\ev{V}}{N^\alpha\Sigma_V}\rightarrow X,
	\label{eq:clt}
\end{equation}
where $X$ is some random variable and "$\rightarrow$" means convergence in distribution for $N\rightarrow\infty$. In the language of FS the same equation reads
\begin{equation}
	\frac{\sum_{n=1}^NV_n-\ev{f}}{K\ev{f}^\alpha}\rightarrow X,
\end{equation}
where $K$ is a constant. The conceptual difference is only that since we know that $\ev{f}=N\ev{V}$, we can use $\ev{f}\ev{V}^{-1}$ as a surrogate variable for $N$.

This analogy tells us that the appearance of FS throughout disciplines might be due to the generality of certain limit theorems. The trivial example is of course that of i.i.d. variables with positive mean and finite variance, leading to the value $\alpha = 1/2$, but there are several other cases. 

If the $V$'s are i.i.d., but their distribution decays asymptotically as $\mathbb P(f) \propto f^{-(\lambda+1)}$ with $0 < \lambda < 2$, then the L\'evy-Gnedenko central limit theorem \footnote{In fact the conditions of the L\'evy-Gnedenko central limit theorem are somewhat looser.} is applicable \cite{feller}. That is in spirit similar to Eq. \eqref{eq:clt}, with $\alpha = 1/\lambda$. The difference is that $\sigma$, and if $\lambda \leq 1$ even $\ev{f}$ is infinite. However, for $N < \infty$ they will have some finite effective value, which can show apparent fluctuation scaling with some non-trivial value of $\alpha$.

Contrary to the relative simplicity of independent (and possibly identically distributed) random variables, dependent variables can be extremely diverse. They have no general theory, and the number of universality classes/limit theorems is infinite. Their structure is not always fully described by pairwise correlations and Hurst exponents (cf. Sections \ref{sec:ising}-\ref{sec:binary}). In these cases sometimes there exists no limit distribution, or, e.g., $\alpha < 1/2$ or $\alpha > 1$ in Eq. \eqref{eq:clt} \cite{clt1}.

Even for the usual $1/2 \leq \alpha \leq 1$ values there is only a limited set of results, here we only mention a few inspired by statistical mechanics. In a series of papers Ellis, Newman and Rosen \cite{ellis.newman, ellis.cw, ellis.rosen} show that in some statistical mechanical systems physical quantities can obey Eq. \eqref{eq:clt} with $\alpha = 1-1/2k$, where $k$ is a non-negative integer. For example, in the Curie-Weiss mean-field model the number of up spins obeys $k=2$ and $\alpha = 3/4$ at criticality, and the distribution of $X$ can also be given explicitly. In Section \ref{sec:ising} we arrived at the same exponent using heuristic arguments. Baldovin and Stella \cite{baldovin.clt} recently published some more general results on a mean-field theory of strongly correlated random variables. In their model fine-tuning the strength of correlations allows for any $1/2\leq\alpha\leq 1$, much in the spirit of Section \ref{sec:binary} and Ref. \cite{ballantyne.correls}.

\subsection{The connection of ensemble and temporal averages}
\label{toc:5.3}

Let us now return to the connection between the TFS 
\begin{equation*}
	\sigma_i(\Delta t) \propto \ev{f_i}^{\alpha_T}, \tag{\ref{eq:taylori}}
\end{equation*}
and the EFS
\begin{equation*}
	\evl{\sigma_N}(\Delta t) \propto {\overline{f_N}}^{\alpha_E} \tag{\ref{eq:taylorN}}
\end{equation*}
laws. These correspond to two definitions of the statistical quantities: 
(i) for Eq. \eqref{eq:taylori} the mean and the standard deviation are calculated as temporal averages; (ii) for Eq. \eqref{eq:taylorN} they are calculated on an ensemble of subsystems of the same size.

For the mere existence of such quantities it is necessary to assume that: (i) signals with the same mean have the same statistical properties, and the processes are \emph{stationary}; or (ii) systems of the same size can be considered elements of the same statistical ensemble, which is a kind of a \emph{homogeneity} condition. In real systems neither of these conditions holds exactly, but they often prove to be good approximations. A deviation from these assumptions is one possible source of the observed broadening of the scaling plots. For example, in the case of precipitation data in Section \ref{sec:precipitation} we found that mean precipitation is not the only determinant of the amplitude of fluctuations. Areas with the same mean precipitation are not equivalent, because they can correspond to very different climates. Factors such as height and geographical position are also relevant.

A related concern is the presence of correlations \cite{gaston.mcardle}. The observations may be correlated in space or time. (i) Two nodes (e.g., populations or weather stations) which are located close to each other can have significant cross-correlations. Fits can be biased, because the observations are not independent. (ii) The signals of individual nodes can have strong temporal autocorrelations, which can amplify statistical errors when the time series are not long enough \cite{smallsampletaylor}.

A more delicate question is the connection between the two types of FS, which has been very vaguely investigated in real systems so far. First of all, the two factors cannot be separated completely. McArdle et al. \cite{mcardle.variation} point out this problem through the example of animal populations. (i) The measurement of the number of individuals in an area takes a finite time. There is an in and outflow of individuals, so the number fluctuates. Thus temporal dynamics can affect the results. (ii) If we want to measure the time series of the size of a given population, we have to assign a spatial scale as what to consider a population. The temporal dynamics will depend on this spatial scale of sampling, possibly in a non-trivial way.

Taylor and Woiwod \cite{taylor.woiwod} conducted a very large scale study of the two (temporal and ensemble) FS laws in animal populations. A systematic comparison is possible when the same sites sampled are at the same time \cite{mcardle.variation}. Taylor and Woiwod calculated the temporal and spatial means and standard deviations of the abundance of some aphids, moths and bird, then calculated $\alpha_\mathrm{T}$ and $\alpha_\mathrm{E}$ for each species. 

First of all, they found that the temporal and ensemble \emph{means} of population differ significantly. Thus it is not surprising that the values of $\alpha$ differ as well. There was absolutely no systematic relationship between $\alpha_\mathrm{T}$ and $\alpha_\mathrm{E}$, and even the same species can show several such values depending on its natural environment. Rather interestingly, the only systematic dependence between species is the presence of positive correlations between the value of $\alpha$'s, and average population size. For example, for temporal data this means that across species $\alpha_\mathrm{T}$ is correlated with $1/M\sum_{i=1}^M \ev{f_i}$. The correlations are present in both cases, although stronger for the temporal variant. Taylor and Woiwod \cite{taylor.woiwod} also suggested, that the interactions between individuals might contribute to the ensemble law more than to the temporal one.

In some studies such as ecology or climatology the definition of the spatial scale comes naturally. Still, most systems have some hierarchical structure on which a degree of aggregation is possible. For example, it is possible to analyze the fluctuations of Internet traffic at the autonomous system level instead of the router level, which might have a different dynamics. On the stock market TFS holds not only for individual stocks, but also when we consider the trading activity of industry sectors \cite{uponrequest}.

In sum, the relationship between ensemble and temporal fluctuation scaling is rather unclear in real systems. The exponents $\alpha_\mathrm{E}$ and $\alpha_\mathrm{T}$ are seldom calculated for the same system, and when they are calculated, they have different values.

\subsection{Fluctuation scaling for growth rates}
\label{toc:5.4}

To be able to interpret FS for temporal fluctuations one has to assume that the underlying system is stationary. For example, in the binary forest model of Section \ref{sec:binary} we assumed that the number of trees is constant and we neglect the contribution of reproduction to the population. To depart from stationarity, we can consider a growing population of $N_i(t)$ individuals, all of which can reproduce at a time $t$ [$V_{i,n}(t)=$ the number of offsprings], die [$V_{i,n}(t)=-1$] or do nothing [$V_{i,n}(t)=0$]. The population can now be written as a sum $$N_i(t+1) = N_i(t) + \sum_{n=1}^{N_i(t)} V_{i,n}(t).$$ $N_i(t)$ is obviously not stationary, because its distribution depends on its value in the previous time step. Nevertheless, one can still construct something similar to TFS by using a restricted ensemble average as follows.

Let us define the growth rate of a population as $$f_i(t)=N_i(t+1)-N_i(t).$$ Now let us make an ensemble of growth observations when the initial population was $N$. The growth rate in this restricted sample is given by $$f_N = \sum_{n=1}^N V_{n}.$$ Because $f$ can be negative, to postulate TFS for size dependence it is more convenient to use $N$ and not $\ev{f}$ as the scaling variable. We conjecture
\begin{equation}
	\sigma_N = \sqrt{\ev{f_N^2}-\ev{f_N}^2} \propto N^\alpha
	\label{eq:taylorgrowth}
\end{equation}
in the spirit of the previous sections, and this is exactly what is found in many systems.

Keitt and Stanley show \cite{keitt.pop, keitt.scaling} that the growth rate fluctuations of animal populations scale as a non-trivial power of the initial population $N$. The finding is not specific for population growth, but occurs in many settings where a positive quantity fluctuates by the addition and subtraction of increments. The same behavior was found by Lee et al. \cite{stanley.firm} for the growth rates of business firms, and Amaral et al. \cite{amaral.growth} even presents a model of the complex structure of the business growth process which predicts the correct exponent. 

J\'anosi and Gallas \cite{janosi.danube} criticize these results, and show the same distribution of growth rates and the scaling law \eqref{eq:taylorgrowth} for the water level fluctuations of the river Danube, which trivially must have a structure that is very different from business firms. Moreover, they show that the daily absolute change of water level scales with the average water level on the same day, and there is a clear crossover behavior between two scaling regimes with $\alpha = 1/2$ and $\alpha = 1$. A related study by Dahlstedt and Jensen \cite{dahlstedt.river} estimates $\alpha \approx 0.9-1$, and suggests that FS can be decomposed into two distinct scaling laws: $\sigma_A \propto A^a$ and $\ev{f_A} \propto A^b$, where $A$ is the area of the river basin.

From the above it is clear that the size-dependent scaling of growth rate fluctuations is a variant of fluctuation scaling for nonstationary (growing) populations. The same formalism can be applied in both cases, and many results could be mutually applied.

\section{Conclusions}
\label{sec:conc}
\label{toc:6}
The aim of this review was to provide a broader perspective on Taylor's law and fluctuation scaling, and to encourage the collaboration between disciplines where these phenomenona are observed. We also outlined a classification scheme on the meaning of the FS exponents. The main conclusion is that several types of mechanisms can lead to the same value of $\alpha$. A similar concern was formulated in the 1982 paper of Taylor and Woiwod \cite{taylor.woiwod}: 

"Extrapolation of dynamic principles from [...] observation is likely to be misleading. We find great differences between [species], but the overlap is also very large. Whilst it is improbable that the details of [...] behaviour in a bird and an aphid would be alike, there are common elements in the [...] structure of their populations." 

While fluctuation scaling alone is not enough to \emph{identify} the underlying dynamics of a system, it is useful for \emph{excluding} some possibilities, and for rejecting certain models which would generate unrealistic $\alpha$'s. Empirical data from virtually all fields of science display fluctuation scaling, and so it is possible to make statements about almost any system where such data is available. For this very reason, in order to deepen our understanding of the phenomenon, it is becoming increasingly important to bridge the gap between several disciplines. But the most puzzling question still remains: Why do, for example, email traffic, stock market trading and the printing activity in a computer lab behave in similar, non-trivial ways? Some insights can be gained from the time window dependence of $\alpha$. That can reveal whether on some time scale the behavior of the system reduces to something simpler or possibly trivial. One can also make efforts to decompose the signals into well-defined constituents, so that a mean-field model based on sums of random variables can be applied. We believe that a possible common origin of all fluctuation scaling laws is the generality of these underlying mean-field type mathematical structures.

\section*{Acknowledgments}

Writing this manuscript would not have been possible without the help of a lot of people. The authors thank P\'eter Csermely for advice on Taylor's law in ecology. They thank B\'alint T\'oth for discussions of statistical physics and limit theorems. They thank Jari Saram\"aki for his comments and Jukka-Pekka Onnela for a critical reading and countless useful remarks. They also thank Maya Paczuski, Peter Grassberger and Albert-L\'aszl\'o Barab\'asi for their ideas on fluctuation scaling. They are grateful to Walter Koenig for data on the reproductive activity of trees, Ricardo Azevedo for cell count data and Jordi Duch and Alex Arenas for their Internet dataset. Ford Ballantyne IV, Marm Kilpatrick and Joe N. Perry are acknowledged for their help with the sections on population dynamics. The authors also thankfully acknowledge correspondence with Jayanth R. Banavar and Andrea Rinaldo on scaling laws in ecology. Finally, they thank Imre J\'anosi for his help and criticism on the analysis of precipitation data. ZE is grateful to Jean-Philippe Bouchaud and for the hospitality of l'Ecole de Physique des Houches. This work was supported by OTKA K60456 and OTKA T049238.

\appendix

\section{The components of the fluctuation $\sigma^2$}
\label{app:components}
\label{toc:A}

A large part of this review is concerned with the standard deviation of the sums of random variables.
This is defined as
$$\sigma^2=\ev{\left (\sum_{n=1}^{N} V_{n}\right)^2}-\ev{\sum_{n=1}^{N} V_{n}}^2,$$
where $V_{n}$ are the individual (not necessarily independent) random variables, and $N$ is the number of these variables which itself can be random.

Let $\mathbb P(N)$ be the probability that the number of variables is $N$. The sum of $N$ variables can be written as $V_N=\sum_{n=1}^{N} V_n$. Let $\mathbb P(V_N)$ denote the density function of this sum when $N$ is fixed. Then the standard deviation of the sum when $N$ itself is a random variable is

\begin{eqnarray}
\sigma^2=\sum_N \mathbb P(N)\int dV_N\mathbb P(V_N)V_N^2 - \nonumber \\ 
\left (\sum_N \mathbb P(N)\int dV_N\mathbb P(V_N)V_N\right )^2 = \nonumber \\
\sum_N\mathbb P(N)\underbrace{\left [\underbrace{\int dV_N\mathbb P(V_N)V_N^2}_{\ev{V^2_N}}- \underbrace{{\left (\int dV_N\mathbb P(V_N)V_N\right )^2}}_{\ev{V_N}^2}\right ]}_{\Sigma^2_{V_N}=N\Sigma^2_V}+ \nonumber \\
\sum_N\mathbb P(N)\underbrace{{\left (\int dV_N\mathbb P(V_N)V_N\right )^2}}_{\ev{V_N}^2=N^2\ev{V}^2}- \nonumber \\
\left (\sum_N \mathbb P(N)\underbrace{\int dV_N\mathbb P(V_N)V_N}_{\ev{V_N}=N\ev{V}}\right )^2= \Sigma^2_V\underbrace{\sum_N\mathbb P(N)N}_{\ev{N}}+\nonumber
\end{eqnarray}
\begin{eqnarray} 
\ev{V}^2\underbrace{\left [\underbrace{\sum_N\mathbb P(N)N^2}_{\ev{N^2}}-\underbrace{\left (\sum_N\mathbb P(N)N\right )^2}_{\ev{N}^2}\right ]}_{\Sigma^2_N} \nonumber
\end{eqnarray}
Thus finally
$$\sigma^2=\Sigma^2_V\ev{N}+\ev{V}^2\Sigma^2_N.$$
In the case when the $V_n$'s are strongly (i.e., power law) correlated $\Sigma^2_{V_N}=\Sigma^2_V N^{2H_V}$ where $H_V$ is the Hurst exponent as defined in Eq. \eqref{eq:tickhurst}, and so
$$\sigma^2=\Sigma^2_V\ev{N^{2H_V}}+\ev{V}^2\Sigma^2_N.$$
The correlations in $N$ are not reflected directly in this expression. Instead, they affect how $\Sigma_N$ changes with the time window size $\Delta t$ as pointed out in Section \ref{sec:gamma}.

\section{Tweedie models and impact inhomogeneity}
\label{toc:B}
\label{app:tweedie}
In this appendix we prove that origin of the non-trivial $\alpha$ values in the formalism of Kendal \cite{kendal.ecological, kendal.blood} is essentially due to impact inhomogeneity. Kendal's formalism is based on the so-called Tweedie exponential dispersion models \cite{penis}. These form a family of random distributions, characterized by the logarithmic cumulant function (see Ref. \cite{penis}, p. 1516)
\begin{equation}
	K_f^*(s) = \ln \ev{e^{sf}}_f = \lambda_\theta [g_\theta(s)-1],
	\label{eq:tweedie.cum}
\end{equation}
where $s$ is a constant and $f$ is the random variable. We use natural logarithms ($\ln$), as opposed to other parts of this review, where we used $10$-base logarithms ($\log$). We also introduced the notation $\ev{x}_y = \int_0^\infty dye^{xy}$. The two terms above are
\begin{equation}
	\lambda = \frac{a - 1}{ka}\left(\frac{k\theta}{1-a}\right)^a,
	\label{eq:tweedie.lambda}
\end{equation}
and
\begin{equation}
	g_\theta(s)=\left (1+\frac{s}{\theta}\right )^a.
	\label{eq:tweedie.g}
\end{equation}
$\theta > 0$ and $a<0$ 

As pointed out both by Kendal \cite{kendal.ecological} and Bar-Lev and Enis \cite{penis}, the form \eqref{eq:tweedie.cum} is characteristic of compound Poisson processes \cite{feller}. These are distributions of random variables of the following type:
\begin{equation}
	f = \sum_{n=1}^{N} V_n, \nonumber
\end{equation}
where $N$ is Poisson distributed, and $V_n$ are i.i.d. random variables. The proof is straightforward, however, we include it here for completeness. The density function of a compound Poisson variable $f$ can be written as a complete probability
$$P(f) = \sum_{N=0}^\infty P(f\vert N)P(N).$$
The characteristic function is given by
$$\ev{e^{sf}}_f=\int df e^{sf}P(f)=$$
$$\sum_{N=0}^\infty P(N)\int df e^{sf} P(f\vert N)=$$
$$\sum_{N=0}^\infty P(N)\int df e^{sf} P(V_1+V_2+\dots V_N)=$$
$$\sum_{N=0}^\infty P(N)\ev{e^{sV}}_V^N.$$
For the last equality we used the property of the characteristic function that $\ev{e^{s\sum_n V_N}}=\ev{e^{sV}}^N.$ Then, knowing that if $N$ is Poisson distributed with mean $\ev{N}$ then its characteristic function is $\ev{e^{tN}}=e^{\ev{N}(e^t-1)}$,
\begin{eqnarray}
\sum_{N=0}^\infty P(N)\ev{e^{sV}}_V^N = \ev{e^{N\ln \ev{\exp(sV)}_V}}_N=\nonumber \\
e^{\ev{N}(\ev{\exp(sV)}_V-1)} = \ev{e^{sf}}_f.
\label{eq:tweedie.calc}
\end{eqnarray}
The next step is to compare Eqs. \eqref{eq:tweedie.lambda}, \eqref{eq:tweedie.g} and \eqref{eq:tweedie.calc}. One finds that for the Tweedie model
\begin{equation}
	\ev{e^{sV}}_V = g_\theta(s) = \left (1+\frac{s}{\theta}\right )^a.\nonumber
\end{equation}
This is the characteristic function of a gamma distribution. Its moments can be determined as usual:
\begin{equation}
	\ev{V} = \left [\frac{\partial}{\partial s} \ev{e^{sV}}_V \right ]_{s=0}=a\theta^{-1},\nonumber
\end{equation}
\begin{equation}
	\ev{V^2} = \left [\frac{\partial^2}{\partial s^2} \ev{e^{sV}}_V \right ]_{s=0}=a(a - 1)\theta^{-2},\nonumber
\end{equation}
\begin{equation}
	\Sigma^2_V = \ev{V^2}-\ev{V}^2 = -a\theta^{-2}.\nonumber
\end{equation}
For the expectation value of the Poisson variable:
\begin{equation}
	\ev{N} = \lambda = \frac{a - 1}{ka}\left(\frac{k\theta}{a-1}\right)^a \propto \theta^a.\nonumber
\end{equation}
Furthermore,
\begin{equation}
	\ev{f} = \left [\frac{\partial}{\partial s} K_f^*(s) \right ]_{s=0} = \lambda a \theta^{-1},\nonumber
\end{equation}
and 
\begin{equation}
	\sigma^2 = \left [\frac{\partial^2}{\partial s^2} K_f^*(s) \right ]_{s=0} = \lambda a(a-1) \theta^{-2}.\nonumber
\end{equation}
Let us recall, that also $\lambda$ contains terms with $\theta$. Simple calculation yields
\begin{equation}
	\sigma^2 = k\ev{f}^{2\alpha},\nonumber
\end{equation}
where $2\alpha = (a-2)/(a-1)$. Consequently:
\begin{enumerate}[(i)]
	\item Let us introduce $\beta = -1/a$. Then 
		\begin{equation*}
			\ev{V} \propto \ev{N}^\beta, \tag{\ref{eq:beta}}
		\end{equation*}
 		which is impact inhomogeneity.
	\item Moreover, $\sigma \propto \ev{f}^\alpha$ with 
		\begin{equation*}
			\alpha = \frac{1}{2}\left(1+\frac{\beta}{\beta+1}\right), \tag{\ref{eq:alphabeta}}
		\end{equation*}
		exactly the same relationship as in Section \ref{sec:beta}.
\end{enumerate}

\section{Fluctuations in the network random walker model}
\label{app:rwm}
\label{toc:C}

This section contains calculations starting from the master equation \eqref{eq:master}. The total number of visitations to node $i$ is the sum over all steps. The substitution of Eq. \eqref{eq:master} gives
\begin{equation}
	f_i = \sum_{s=1}^{s_\mathrm{max}} N_i(s) = \sum_{s=0}^{s_\mathrm{max}-1} \sum_{j\in \mathcal K_i}\sum_{n=1}^{N_j(t-1)} \delta_{n}(j\rightarrow i; s),
\end{equation}
where $\delta_{n,s}(j\rightarrow i; s)$ is a variable which is $1$ if the $n$'th token of node $j$ in step $s$ jumps to node $i$ (happens with probability $1/k_j$), and $0$ otherwise. $k_i$ is the degree of node $i$, $\mathcal K_i$ is the set of neighbors of node $i$, and $N_j(t=0)$ corresponds to the initial condition. 

For any finite network one can switch the order of the first two sums, and in
$$f_i = \sum_{s=1}^{s_\mathrm{max}} N_i(s) = \sum_{j\in \mathcal K_i}\sum_{s=0}^{s_\mathrm{max}-1}\sum_{n=1}^{N_j(t-1)} \delta_{n}(j\rightarrow i;s)$$
if $S$ is taken large and because for any fixed $n$ the variables $\delta_{n,s}(j\rightarrow i)$ are independent, then due to the central limit theorem the last two sums converge to independent Gaussians:
\begin{equation}
f_i = \sum_{s=1}^{s_\mathrm{max}} N_i(s) = \sum_{j\in \mathcal K_i}\left (\frac{s_\mathrm{max}\ev{N_j}}{k_j}+\sqrt{\frac{s_\mathrm{max}\ev{N_j}}{k_j}}{\mathcal G}_j(s)\right ),
\label{eq:eqf}
\end{equation}
where ${\mathcal G}_j(s)$ are i.i.d. standard Gaussians such that
\begin{equation}
	\ev{{\mathcal G}_i(s){\mathcal G}_j(r)} = \delta_{ij}\delta_{sr},
	\label{eq:gausscc}
\end{equation}
where the right hand side has two Kronecker-deltas.
Consequently
\begin{equation}
	\ev{f_i(t)f_j(t)}=\ev{f_i(t)}\ev{f_j(t)},\ \mathrm{when}\ i\not = j.
	\label{eq:fcc}
\end{equation}

One can take the expectation value of the left hand side of Eq. \eqref{eq:eqf}. Finally,
\begin{equation}
\ev{f_i} = \sum_{j\in \mathcal K_i}\frac{s_\mathrm{max}\ev{N_j}}{k_j}.
\label{eq:meanf0}
\end{equation}
By substitution one can check that the solution is
\begin{equation}
\ev{f_i} = s_\mathrm{max}\ev{N_i}=k_i\frac{s_\mathrm{max}W}{\sum_j k_j},
\label{eq:meanf}
\end{equation}
and all the walkers are accounted for: $\sum_i \ev{f_i} = s_\mathrm{max}W$.

Now let us calculate the standard deviation for both sides of Eq. \eqref{eq:eqf}:
\begin{eqnarray*}
\sigma_i^2 = \ev{\left [\sum_{j\in \mathcal K_i}\left (\frac{s_\mathrm{max}\ev{N_j}}{k_j}+\sqrt{\frac{s_\mathrm{max}\ev{N_j}}{k_j}}{\mathcal G}_j\right )\right ]^2} - \nonumber \\
\ev{\sum_{j\in \mathcal K_i}\underbrace{\left (\frac{s_\mathrm{max}\ev{N_j}}{k_j}+\sqrt{\frac{s_\mathrm{max}\ev{N_j}}{k_j}}{\mathcal G}_j\right )}_{(a)}}^2 = \dots
\end{eqnarray*}
$(a)$ can be replaced by $\frac{s_\mathrm{max}\ev{N_j}}{k_j}$, because $\ev{{\mathcal G}_j}=0$.
\begin{eqnarray*}
\sigma^2_i = \underbrace{\ev{\left (\sum_{j\in \mathcal K_i}\frac{s_\mathrm{max}\ev{N_j}}{k_j}\right )^2}}_{(b)} + \nonumber
\end{eqnarray*} 
\begin{eqnarray*}
\underbrace{2\ev{\left (\sum_{j\in \mathcal K_i}\frac{s_\mathrm{max}\ev{N_j}}{k_j}\right )\left (\sum_{l\in \mathcal K_i}\sqrt{\frac{s_\mathrm{max}\ev{N_l}}{k_l}}{\mathcal G}_l\right )}}_{(c)} + \nonumber \\ \underbrace{\ev{\left (\sum_{l\in \mathcal K_i}\sqrt{\frac{s_\mathrm{max}\ev{N_l}}{k_l}}{\mathcal G}_l\right )^2}}_{(d)} - \underbrace{\ev{\sum_{j\in \mathcal K_i}\frac{s_\mathrm{max}\ev{N_j}}{k_j}}^2}_{(e)}
\label{eq:eqsigma2}
\end{eqnarray*}
One can use Eq. \eqref{eq:fcc} to write
\begin{equation*}
(b) = \ev{\sum_{j\in \mathcal K_i}\frac{s_\mathrm{max}\ev{N_j}^2}{k_j^2}} + \sum_{j\not = l \in \mathcal K_i}\frac{s_\mathrm{max}^2\ev{N_j}\ev{N_l}}{k_jk_l}.
\end{equation*}
$(c) = 0$, because of Eq. $\ev{{\mathcal G}_l}=0$.
\begin{equation*}
(d) = \ev{\sum_{l\in \mathcal K_i}\left (\sqrt{\frac{s_\mathrm{max}\ev{N_l}}{k_l}}{\mathcal G}_l\right )^2} = \ev{\sum_{l\in \mathcal K_i}\frac{s_\mathrm{max}\ev{N_l}}{k_l}},
\end{equation*}
because of \eqref{eq:gausscc}. By changing a summation variable, one can write
\begin{equation}
(e) = \sum_{j,l\in \mathcal K_i}\frac{s_\mathrm{max}^2\ev{N_j}\ev{N_l}}{k_jk_l}
\end{equation}
Combining all the above, one gets
\begin{eqnarray*}
\underbrace{\ev{\sum_{j\in \mathcal K_i}\frac{s_\mathrm{max}\ev{N_j}^2}{k_j^2}}}_{(f)} + \underbrace{\sum_{j\not = l \in \mathcal K_i}\frac{s_\mathrm{max}^2\ev{N_j}\ev{N_l}}{k_jk_l}}_{(g)} + \nonumber \\ \sum_{l\in \mathcal K_i}\frac{s_\mathrm{max}\ev{N_l}}{k_l}- \underbrace{\sum_{j,l\in \mathcal K_i}\frac{s_\mathrm{max}^2\ev{N_j}\ev{N_l}}{k_jk_l}}_{(h)}.
\end{eqnarray*}
$(f)$ and $(g)$ and $(h)$ combine to 
\begin{eqnarray*}
\ev{\sum_{j\in \mathcal K_i}\frac{s_\mathrm{max}\ev{N_j}^2}{k_j^2}} - \sum_{j\in \mathcal K_i}\frac{s_\mathrm{max}\ev{N_j}^2}{k_j^2} \equiv \sum_{j\in \mathcal K_i} \frac{\sigma^2_j}{k_j}.
\end{eqnarray*}
Then,
\begin{equation*}
\sigma_i^2 = \sum_{j\in \mathcal K_i} \frac{\sigma_j^2}{k_j^2}+\sum_{j\in \mathcal K_i} \frac{s_\mathrm{max}\ev{N_j}}{k_j}.
\end{equation*}
The second term can be evaluated from Eq. \eqref{eq:meanf0}, to find
\begin{equation}
\sigma_i^2 = \sum_{j\in \mathcal K_i} \frac{\sigma_j^2}{k_j^2}+\ev{f_i}.
\end{equation}

\bibliographystyle{prsty}
\bibliography{taylorism}

\end{document}